\documentclass[preprint2]{emulateapj}
\usepackage{slashbox}
\usepackage{rotating}
\usepackage{textcomp}
\usepackage{amsmath}
\usepackage{natbib}
\usepackage{Sweave}
\usepackage{hyperref}
\usepackage{csvsimple}
\newcommand\totnum{2214}
\newcommand\completenum{54}

\begin{document}

\title{A Comprehensive Statistical Description of Radio-Through-$\gamma$-Ray
Spectral Energy Distributions of All Known Blazars}

\author{Peiyuan Mao$^{1}$, C. Megan Urry$^{1}$, 
Francesco Massaro$^{1, 4, 5}$, Alessandro Paggi$^{2}$,
Joe Cauteruccio$^{3}$ and Soren R. K\"unzel$^{3}$} 

\affil{
$^{1}$ Yale Center for Astronomy \& Astrophysics, Physics Department, New Haven, CT 06520 \\
$^{2}$ Harvard-Smithsonian Center for Astrophysics, Cambridge, MA 02138 \\
$^{3}$ Department of Statistics, Yale University, New Haven, CT 06520 \\
$^{4}$ Physics Department, University of Turin, via Pietro Giuria 1, I-10125 Turin, Italy \\
$^{5}$ Istituto Nazionale di Fisica Nucleare, Sezione di Torino, I-10125 Turin, Italy
}

\begin{abstract}
    We combined multi-wavelength data for blazars 
    from the Roma-BZCAT
    catalog and analyzed 
    hundreds of X-ray spectra. 
    We present the fluxes and Spectral Energy Distributions (SEDs), 
    in 12 frequency bands from radio to $\gamma$-rays, for a 
    final sample of \totnum\ blazars.
    Using a model-independent statistical approach, we 
    looked for systematic trends in the 
    SEDs; the most significant trends involved the  
    radio luminosities and X-ray spectral indices of the blazars.
    We used a Principal Component Analysis (PCA),
    to determine the basis vectors of the blazar
    SEDs and, in order to maximize the size of the sample, imputed 
    missing fluxes using the K-nearest neighbors method.

    Using more than an order of magnitude more data than was available 
    when Fossati et al. (1997, 1998) first reported trends of SED shape with blazar luminosity, 
    we confirmed the anti-correlation 
    between radio luminosity and synchrotron peak frequency, 
    although with greater scatter than was seen in the smaller sample.
    The same trend can be seen between bolometric luminosity 
    and synchrotron peak frequency.
    Finally, we used all available blazar data 
    to determine an empirical SED description that depends
    only on the radio luminosity at 1.4~GHz and the redshift. 
    We verified that this statistically significant relation 
    was not a result of the luminosity-luminosity correlations that are natural in flux-limited 
    samples (i.e., where the correlation is actually caused 
    by the redshift rather than the luminosity).

\end{abstract}


\section{Introduction} 
\label{sec:introduction}

Blazars are a class of active galactic nuclei (AGN) marked by 
large amplitude and rapid variability, superluminal motion, 
and strong non-thermal emission across the entire electromagnetic spectrum
from radio to $\gamma$-rays
\citep{urry95}.
Their spectral energy distributions (SEDs) are dominated by emission from
a Doppler-boosted relativistic jet angled close to the line of sight \citep{blanford78, urry82}. 
Two distinct blazar subclasses 
have been defined:
flat-spectrum radio quasars (FSRQs), 
which show strong, broad emission lines in their optical-IR spectra, 
and BL Lac objects, which have weak or no emission lines (equivalent widths EW$<$ 5\AA;
\citealp{angel80, stocke85, landt01}).

Blazar SEDs are characterized by two broad components, 
the low frequency one produced by synchrotron radiation 
from relativistic electrons in the jet, 
and the high frequency one produced by inverse Compton scattering 
of ambient photons by those same electrons
\citep{urry95, abdo10}.
Blazars with synchrotron peaks at high frequency 
(HSPs, with $\nu_{peak}>10^{15}$Hz, see \citealp{abdo10})
have been found primarily in X-ray surveys, while
low-synchrotron-peaked blazars (LSPs, with $\nu_{peak}<10^{14}$Hz) 
have been found preferentially in radio and GeV $\gamma$-ray surveys
\citep{giommi94,stocke85,maraschi86,fichtel94}.
In the early blazar samples, the synchrotron peak frequencies were anti-correlated 
with luminosity 
\citep{stocke85, sambruna96, fossati97, giommi94}, 
However, these studies were based no more than 
$\sim130$ 
objects,
with at most 34 \citep{fossati98} having $\gamma$-ray data,
so the full range of luminosity and redshift was not well sampled.

With the launch of \emph{Fermi} satellite \citep{ackermann11},
GeV $\gamma$-ray data are now available for more than 1000 
blazars, along with extensive multi-wavelength data at radio, infrared, optical 
and X-ray wavelengths.
In this paper, we characterize the SEDs of \totnum\ blazars for which redshifts 
and 
substantial
wavelength coverage are available, using a non-parametric statistical approach.
While this sample is not complete in a formal sense, it is a far larger sample with far more comprehensive SED data
than most previous studies (\citealp{fossati97, fossati98} cf. \citealp{giommi13}.)
In Section \ref{sec:the_data}, we describe the sample 
and the data collected at various wavelengths. 
Since much of the X-ray data were previously unpublished, we also include a description of the X-ray data reduction. 
In Section \ref{sec:statistical_investigation} we use statistical analysis to find the most significant 
correlations displayed in the data, free from any assumed functional form.
Throughout the paper, the energy spectral index, $\alpha$, is the power-law exponent of the flux density
defined by F$_{\nu}\propto\nu^{-\alpha}$, which makes the photon index $\Gamma = \alpha+1$. 
The cosmological parameters $H_0=70$~km~s$^{-1}$Mpc$^{-1}$ 
and $\Omega_{\Lambda}=0.72$ \citep{hinshaw13} are used. 
All references of the logarithmic function are in base 10.

\section{Sample Selection} 
\label{sec:the_data}

We base our blazar sample on the data in the Multi-wavelength Catalog of 
blazars, Roma-BZCAT\footnote{http://www.asdc.asi.it/bzcat/} v5.0,
released in December 2014,
which is the most comprehensive catalog in the literature, 
\citep{massaro09, massaro11, massaro15} with $3561$ blazars and blazar candidates. 
The 
Roma-BZCAT 
catalog includes: 
$1151$ BL Lac objects (named BZBs in Roma-BZCAT), 
of which $1059$ are spectroscopically confirmed 
(i.e., the spectra show no emission lines with EW$>5$\AA),
and another $92$ are candidates (i.e., they have been classified as BL Lacs 
in the literature but spectroscopic data 
are not in the published literature;
$1909$ radio-loud quasars with flat radio spectra (BZQs);
$227$ blazars of uncertain type (BZUs,
adopted for sources with 
peculiar characteristics\footnote{For instance, occasional presence/absence of broad spectral lines or features, 
transition objects between a radio galaxy and a BL Lac, 
or galaxies hosting a low luminosity blazar nucleus. Refer to http://www.asdc.asi.it/bzcat/ for details});
and $274$ potential BL Lac objects which have optical spectra 
dominated by galaxy emission instead of non-thermal emission 
(BZGs, see \citealp{massaro12}).
Further details about the Roma-BZCAT selection and classification criteria,
together with the list of the surveys on which it is based, 
can be found in \citep{massaro09, massaro11, massaro15}. 

Since the main goal of our study is to investigate the shapes of blazar SEDs,
and to identify their possible dependences on luminosity and redshift, 
we considered only BZBs and BZQs for which a firm redshift\footnote{
    Redshift without the `?', `??', `>' flags, where flags indicate
    uncertainty in the value (`?'), extreme uncertainty (`??') 
    or a lower limit (`>'). Refer to http://www.asdc.asi.it/bzcat/ for details.
}
is known. 
This reduces our sample to $2220$ sources, 
including $1880$ BZQs and $340$ BZBs 
(not surprisingly, only a third of the $1059$ BZBs have redshifts);
the $92$ BL Lac candidates are not included
because
we want a clean sample not contaminated by sources 
of uncertain type.
Cross-matching with other multi-wavelength surveys,
all $2220$ were detected in the radio band (that being one of their defining characteristics), 
most were detected in the near-IR, slightly
fewer than half have counterparts in the optical, 
about one third have X-ray data, 
and just under a quarter were detected in $\gamma$-rays.
Below we describe the data used in each wavelength band.

\subsection{Radio Data} 
\label{sub:nvss_radio}
For $1992$ blazars, radio flux densities at 1.4~GHz are available from the 
NRAO VLA Sky Survey (NVSS) \citep{condon98}
and/or the Faint Images of the Radio Sky at Twenty Centimeters (FIRST) survey
\citep{becker95, white97}.
The NVSS surveyed the 82\% of the celestial sphere
at declinations $\delta \geq -40$\textdegree\,
with a flux limit of $\sim$2.5~mJy and an angular resolution of $\sim 45^{\prime\prime}$,
while FIRST covers 10,000 deg$^2$ in the northern hemisphere
to a sensitivity limit of $\sim$1~mJy with an angular resolution of $\sim 5^{\prime\prime}$.

For another $224$ sources, Roma-BZCAT provides the flux density at $843$ MHz 
from the Sydney University Molonglo Sky Survey (SUMSS) \citep{mauch03},
which covers 3500 deg$^2$ of the southern sky with $\delta\leq-30^{\circ}$
with mosaic resolution of $45''\times 45''$ $cosec |\delta|$ and
a limiting flux of 2-3~mJy, similar to the NVSS.
In order to compare the sources on equal footing, we need a frequency sampling 
as uniform as possible and interpolate these values to 1.4~GHz assuming the power-law spectral index 
measured between 843 MHz
and 5~GHz (the slope of a linear fit between 0.843 and 5~GHz in $\log\nu - \log\nu L_{\nu}$ plane). 
Two of these $224$ sources did not have 5~GHz fluxes reported, 
and another $4$ sources have only 5~GHz fluxes reported;
these $6$ were not included in our sample since 
interpolation to 1.4~GHz was not possible.

\begin{figure}[hbpt]
  \begin{center}
    \includegraphics[width=0.99\columnwidth]{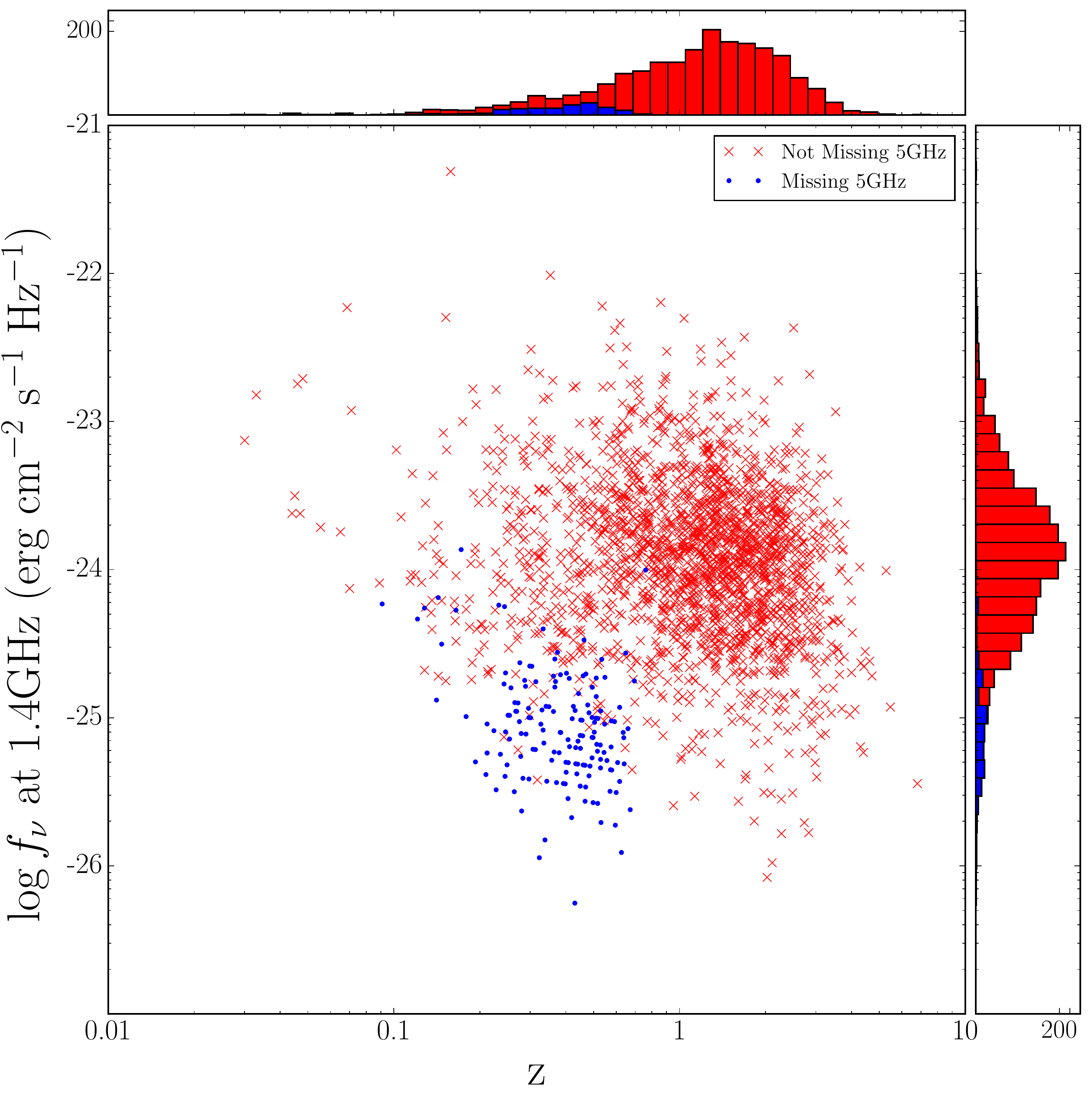}
  \end{center}
  \caption{Radio flux at 1.4~GHz \emph{vs.} redshift for blazars with ({\it red crosses}) 
or without ({\it blue circles}) 5~GHz data, 
and the associated one-dimensional histograms. The blazars missing 5~GHz fluxes 
are generally at lower redshift 
and lower luminosity than the rest of the sample.
}
\label{fig:missing_radio}
\end{figure}

For all but $158$ of the \totnum\ blazars, 
the radio flux density at $5$~GHz is also available in the Roma-BZCAT 
(the actual band center is at $4.85$~GHz).
The $158$ missing blazars are preferentially 
low redshift objects with 
faint radio fluxes at 1.4~GHz (blue points in Figure~\ref{fig:missing_radio}), 
meaning that either their 5~GHz fluxes are below the NVSS or FIRST sensitivity thresholds,
or they have higher positional uncertainties so the cross-matches between two frequencies are more difficult.
Clearly these are among the lowest luminosity blazars in the sample.

\subsection{Infrared Data} 
\label{sub:infra_red}

The WISE (Wide-field Infrared Survey Explorer) all-sky survey, 
released in November 2013 as the ALLWISE catalog\footnote{http://wise2.ipac.caltech.edu/docs/release/allwise/},
covers the entire sky at 3.4, 4.6, 12, and 22~$\mu$m, 
with angular resolutions of $6.1''$, $6.4''$, $6.5''$ and $12.0''$, respectively.
The $5\sigma$ photometric sensitivity is estimated to be 0.068, 0.098, 0.86 and 5.4~mJy 
(16.6, 15.6, 11.3, 8.0 Vega mag), respectively, in unconfused regions on the ecliptic plane.
We converted the reported WISE magnitudes\footnote{http://wise2.ipac.caltech.edu/docs/release/allwise/expsup/index.html}
to flux densities using zero-magnitude flux densities of 
309.540~Jy (W1), 171.787~Jy (W2), 31.674~Jy (W3) and 8.363~Jy (W4), respectively. 
We also corrected the reported WISE magnitudes for Galactic absorption 
using the \citet{draine03} relation and values of $N_H$ from \citet{kalberla05}; 
and applied the color corrections from \citet[Table 1]{wright10}.

Cross-matching our blazar sample with the WISE survey,
adopting an optimal radius of 3.3\arcsec\ \citep{dabrusco13,massarof13, dabrusco14},
we find that $2151$ out of \totnum\ sources in our sample have a mid-IR counterpart. 
All $2151$ are detected in at least the first two WISE bands 
at 3.4~$\mu$m and 4.6~$\mu$m;
$2019$ were also detected at 12~$\mu$m;
and $1628$ have a counterpart in all 
four filters.
No blazars had more than one WISE counterpart within 3.3\arcsec\ 
and the probability of spurious association of any one WISE source
with a BZCAT source is lower than 3\% \citep{dabrusco13}.

We note that $1478$ sources of the $2151$ with a WISE counterpart also 
have a counterpart in the 2MASS catalog,
but these data were not included in our analysis because the 2MASS data 
have a much brighter flux limit than WISE, 
in part because 
the sky background is so much higher from the ground.

\subsection{Optical Data} 
\label{sub:sloan_optical}

The Sloan Digital Sky Survey (SDSS) 
covers more than 8000 square degrees of the sky 
in five optical band-passes, centered at $354.3$~nm, 
$477$~nm, $623.1$~nm, $762.5$~nm and $914.3$~nm 
(called $u$, $g$, $r$, $i$, $z$; \citealp{stoughton02}).
We used SDSS AB magnitudes from Data Release 9 \citep{ahn12}, 
and converted to flux densities\footnote{http://www.sdss3.org/dr9/algorithms/fluxcal.php} 
using a zero-point flux density of 3631 Jy.
The SDSS covers only about one quarter of the sky 
but more than half the Roma-BZCAT sample
($1238$ objects) lies in its footprint because of an observational bias 
toward the northern hemisphere.

Adopting an association radius of 1.8\arcsec,
which provides a probability of spurious associations lower than 1\%
for a match between any Roma-BZCAT source and an SDSS DR9 source
\citep{massarof14}, 
we find that $963$ of $1238$ blazars have an SDSS counterpart. 
For our work we included only the $432$ SDSS sources that have the following flags\footnote{See description of flags at http://cas.sdss.org/dr7/sp/help/browser/browser.asp?n=PhotoObj.} 
equal to 1: CLASS\_OBJECT (i.e., primary object) 
and CODE\_MISC (i.e., clean photometry for point source).

Although SDSS officially reports fluxes in 5 bands, 
we did not include the u-band 
flux in our analyses, since it is the least sensitive band
and is very likely affected by ``big blue bump'' 
(thermal emission from an optically thick accretion disk 
feeding the central black hole, see \citealt{shields78})
in many blazars, especially FSRQs.
We also excluded magnitudes below the completeness limits
(22.2 for g and r bands, 21.3 for i band and 20.5 for z band\footnote{https://www.sdss3.org/dr9/scope.php}),
as these data tend to be unreliable and 
could greatly affect the other optical 
bands
due to their contribution to caluclating the K corrections in the remaining bands.
Removing them reduces the number of blazars with SDSS counterparts to $407$.
As in the case of WISE cross-matching, no double 
counterparts 
were found.

\subsection{X-Ray Data} 
\label{sub:swift_x_ray}

The Swift X-Ray Telescope (XRT) is a focusing X-ray telescope 
with a $110$~cm$^2$ effective area, $18''$ resolution 
and an energy range $0.2-10$~keV, 
with a flux limit of 
$2\times 10^{-14}$~erg~cm$^{-2}$~s$^{-1}$ 
for a $10^4$-second observation \citep{burrows05}. 

Searching for X-ray counterparts of the \totnum\ blazars in the Swift 
archive\footnote{http://swift.gsfc.nasa.gov/archive/}, 
we found $700$ sources with an X-ray detection above 3$\sigma$.
The Swift sky coverage is far from uniform, with some blazars being observed repeatedly 
and others not having been observed yet; 
observations of previously unobserved blazars have been proposed.

The $34$ X-ray-brightest sources generate pile-ups in the Swift 
detectors (this occurs whenever two or more photons are detected as a single event,
and we take the threshold to be when the count rate exceeds 0.4 counts per second). 
Detailed Swift analyses for these objects were 
reported in two earlier papers 
\citep{massarof08, massarof11}, 
from which we took the normalization at 1~keV and the spectral index.

Another $157$ X-ray sources have count rates $<0.4$ counts per second but a 
substantial number of total counts, i.e., $>400$. 
We analyzed these Swift data with the 
XRT Interactive Tool
provided at the ASDC-BZCAT website,
using the response matrix for the Swift XRT \citet{burrows05}.
The tool runs XSPEC with a power-law model with absorbing column density, $N_H$, 
frozen at the Galactic value (taken from 
http://heasarc.gsfc.nasa.gov/cgi-bin/Tools/w3nh/w3nh.pl;
\citealp{kalberla05}) after binning the data into 6 energy bins,
and yields a best-fit normalization (flux density at $1$~keV) and spectral index. 
As a check, we compare the fitted spectral index to that estimated 
from the measured hardness ratio (discussed below). 

\begin{figure}[hbpt]
  \begin{center}
    \includegraphics[width=1.09\columnwidth]{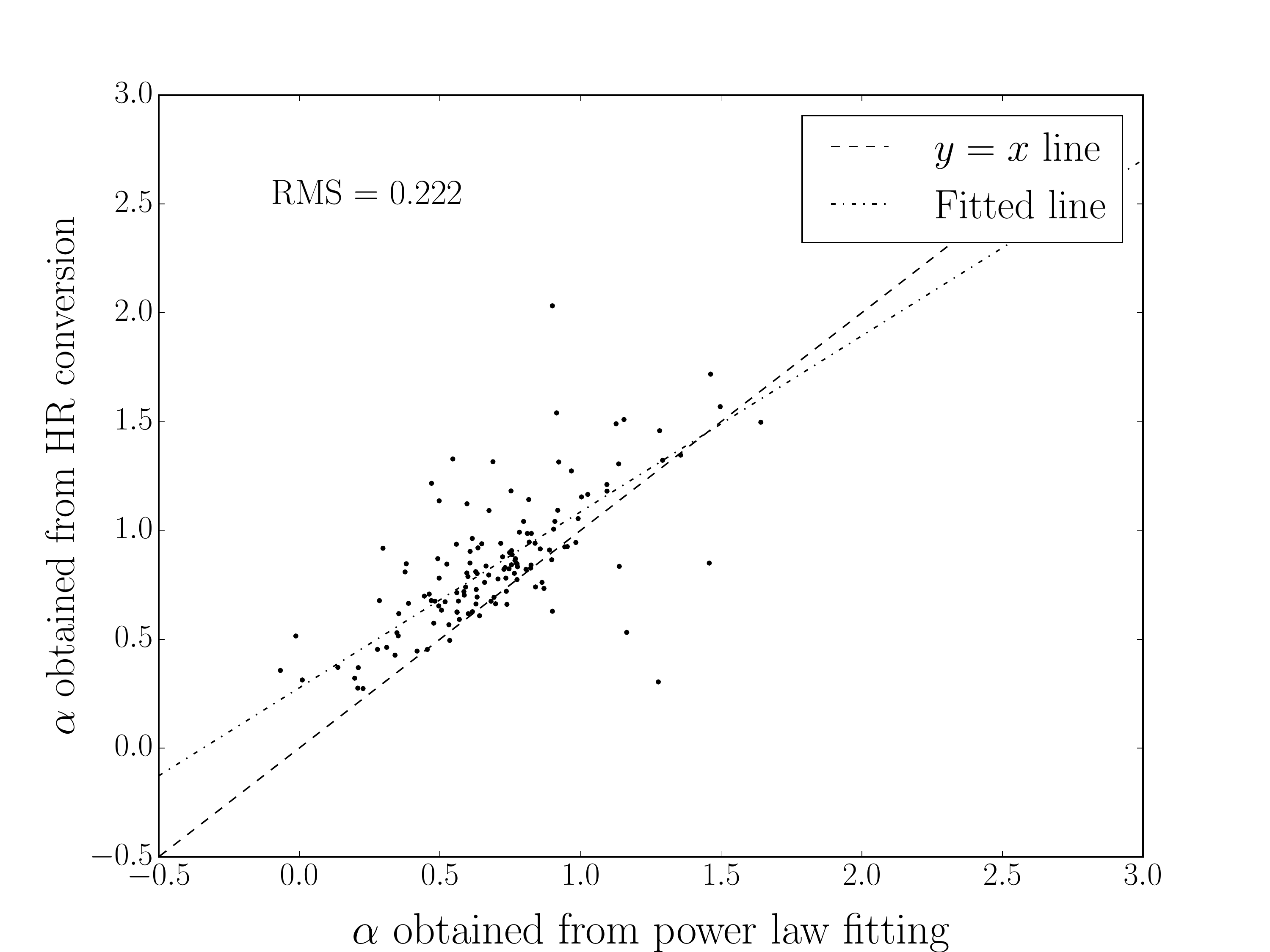}
  \end{center}
  \caption{Comparison of spectral indices obtained
  for moderately bright sources from the hardness ratio (HR) and
from directly fitting a power law with XSPEC. 
The HR-derived index is systematically steeper than the fitted index, but the offset
 is comparable to the rms dispersion ($\Delta\alpha=0.222$) and well within the 
larger uncertainty introduced by the degeneracy of $\alpha$ and $N_H$. 
A correction is applied to all HR-derived spectral indices to correct for 
this difference.}
\label{fig:a_vs_a}
\end{figure}

Fig.~\ref{fig:a_vs_a} shows that the 
spectral index derived from the hardness ratio is systematically steeper 
than the fitted index, by about $\Delta\alpha\sim 0.2$,
comparable to the root mean square (rms) deviation 
and well within the systematic uncertainties. 
A rescaling has been applied to all HR-derived spectral indices
to correct for this difference.

For the remaining $509$ blazars that were too faint for spectral analysis 
(total counts $<$400),
we derived the spectral index from the hardness ratio,
defined as HR $\equiv$ (H-S)/(H+S), where H and S are the counts in 
the hard ($2-10$keV) and soft ($0.5-2$keV) bands, respectively
 \citep{evans14}.
Specifically, we re-ran the Swift XRT pipelines and generated the calibrated
event files and exposure maps for each blazar, 
and measured the number of counts in the soft (0.5-2~keV) and hard (2-10~keV) bands, 
from which we calculated the hardness ratio.
We then used PIMMS \citep{mukai93}, taking into account of the Swift XRT response,
to calculate the flux density at 1~keV using the obtained spectral index
and the total count rate.

\subsection{$\gamma$-Ray Data} 
\label{sub:fermi_gamma_ray}

We obtained $\gamma$-ray data for $571$ of the \totnum\ blazars 
from the \emph{Fermi} Third LAT Sources Catalog
\footnote{http://fermi.gsfc.nasa.gov/ssc/data/access/lat/4yr\_catalog/}
(3FGL, \citealp{acero15}). 
The LAT (Large Area Telescope) is a silicon strip detector sensitive to
$\gamma$-rays in the energy range from 20~MeV to more than 300~GeV \citep{atwood09};
the associations with the Roma-BZCAT were already provided in the 3FGL (3LAC,
the counterpart coordinates were used to find the best cross-matches with the
latest version of the BZCAT). 

The \emph{Fermi} catalog lists integrated fluxes,
in units of photons per unit area per unit time, in 5 energy bands:
100~MeV--300~MeV, 300~MeV--1~GeV, 1~GeV--3~GeV, 3~GeV--10~GeV and 10~GeV--100~GeV.
The best-fit photon index across the entire $\gamma$-ray spectrum, $\Gamma$, 
is also reported there in \citet{acero15}. 
We did not require detections in all 5 bands; 
instead, we derived the monochromatic flux density at 1~GeV
(which is near the peak of the Fermi LAT sensitivity)
from $\Gamma$ and the reported flux (integrated from the 100~MeV to 100~GeV range).

\subsection{Multi-Wavelength Data}

We collected 
available fluxes 
for 
the following 12 frequencies: 
\begin{itemize} \itemsep -2pt
    \item 2 in radio: 1.4~GHz, 5~GHz;
    \item 4 from infra-red (WISE): W1, W2, W3, W4;
    \item 4 from optical (SDSS): g, r, i, z;
    \item 1 in X-ray: 1~keV; and
    \item 1 in $\gamma$-ray: 1~GeV. 
\end{itemize}
All flux densities were converted to the same units, 
erg~cm$^{-2}$~s$^{-1}$~Hz$^{-1}$.
We then applied a K correction using the formula 
$F_{\nu, corr} = F_{\nu}\cdot(1+z)^{\alpha-1}$
where $\alpha$ is the spectral index at frequency $\nu$.
For SDSS and WISE, $\alpha$ comes from fitting a local power law 
between adjacent bands and taking into account the color-color corrections;
for $\gamma$-rays, $\alpha$ comes from the $\Gamma$ values reported 
by \citet{acero15}. 

Note that for 102 blazars the $\gamma$-ray spectrum was fitted as a log parabola
instead of the usual power law, where 
$\frac{dN}{dE} = K( \frac{E}{E_0})^{-\alpha-\beta\log{E/E_0}}$
\citep{acero15}. We computed their fluxes accordingly and applied a modified K correction
consistent with the spectral shape 
($F_{\nu, corr} = F_{\nu}\cdot(1+z)^{1 - \alpha -\beta\log{(1+z)}}$).
K-corrected fluxes were converted to luminosities, 
and to characterize the SED shapes, we fit to $\log\nu L_{\nu}$
(see Section \ref{sec:statistical_investigation}). 

We estimated the frequencies of the synchrotron component
by fitting a second order polynomial to the radio-through-optical SED, 
for the 1590 blazars with radio data and complete optical or complete infrared data. 

\begin{figure}[hbpt]
  \begin{center}
    \includegraphics[width=0.99\columnwidth]{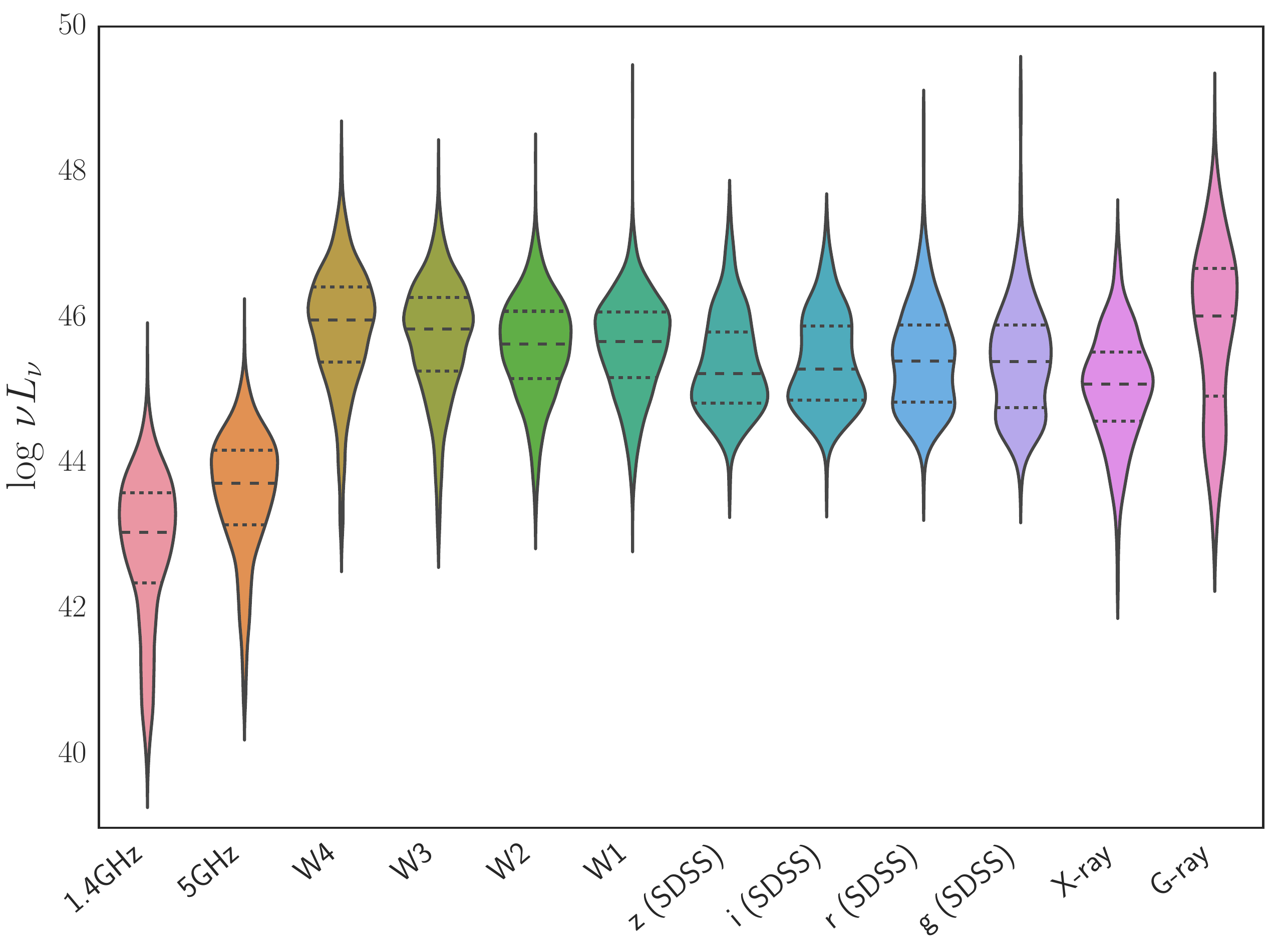}
  \end{center}
  \caption{Violin plot of the blazar sample, showing their luminosity distributions
  ($\log\,\nu L_{\nu}$). The three dotted lines in each ``violin'' shows
  the 25th, 50th and 75th percentiles. Note that each ``violin'' is normalized 
  to the same area, i.e., the shaded area does not represent the relative 
  number of objects in each band. }
\label{fig:violin}
\end{figure}

In summary, our sample includes: 
\totnum\ blazars (1876 BZQs, 338 BZBs; 
this sums to 
less than the radio sample because 
we excluded 6 objects which had fluxes only at 5 GHz),
of which 2151 have WISE data, 407 have SDSS counterparts, 
700 have Swift XRT data,
and 571 have \emph{Fermi} LAT data.
Only \completenum\ blazars have complete data across all frequencies. 
Table \ref{tab:missing} and Figure \ref{fig:violin} summarize 
the multi-wavelength data available for the sample;
the complete data of our blazar collection is in a machine readable format online,
a sample of which is shown in Table \ref{tab:complete}.
\begin{table}[hpbt] 
    \caption{Data Completeness Summary}
\begin{center}
\begin{tabular}{cccc}
    \hline
    Band               
    & Sources\footnote{Number of objects with data available}
    & $L_{min}$\footnote{Minimum value of $\log(\nu L_\nu)$ for the subsample detected in each band, in ergs/s.}    
    & $L_{max}$\footnote{Maximum value of $\log(\nu L_\nu)$ for the subsample detected in each band, in ergs/s.} \\
    \hline
    z                  & \totnum\  &  0.03     & 6.802 \\
    1.4GHz             & \totnum\  &  39.71    & 45.49 \\
    5GHz               & 2056      &  40.57    & 45.89 \\
    W4                 & 1628      &  42.89    & 48.32 \\
    W3                 & 2019      &  42.93    & 48.08 \\
    W2                 & 2151      &  43.14    & 48.20 \\
    W1                 & 2151      &  43.09    & 49.16 \\
    z                  & 407       &  43.66    & 47.47 \\
    i                  & 407       &  43.65    & 47.30 \\
    r                  & 407       &  43.64    & 48.69 \\
    g                  & 407       &  43.64    & 49.12 \\
    X-Ray              & 700       &  42.28    & 47.21 \\
    $\gamma$-Ray       & 571       &  42.89    & 48.70 \\
    \hline
\end{tabular}
\label{tab:missing}
\end{center}
\end{table}

\begin{table*}[hpbt] 
    \caption{The complete data for all blazars \footnote{The full sample table contains 20 columns and \totnum rows, and is available in a machine readable format online.}}
    \small
    \setlength\tabcolsep{2pt}
\begin{center}
\begin{tabular}{ccccccc p{5cm}}
    \hline
    \hline
    BZCAT & WISE & Fermi & R.A. & Dec & Redshift & 1.4~GHz \\
    name & name & name & (degs) & (degs) & &  (erg~s$^{-1}$) \\
    \hline
    BZBJ0945+5757 & J094542.20+575747.6 & 3FGL J0945.9+5756 & 146.425958333 & 57.96325 & 0.229 & 41.2999741162 \\
    BZBJ0002-0024 & J000257.17-002447.2 & N/A & 0.738166666667 & -0.413083333333 & 0.523 & 42.3071693579 \\
    BZBJ0006-0623 & J000613.88-062335.2 & N/A & 1.557875 & -6.39311111111 & 0.347 & 42.9106779635 \\
    BZBJ0008-2339 & J000835.40-233927.8 & 3FGL J0008.6-2340 & 2.14741666667 & -23.6577222222 & 0.147 & 40.3985164399 \\
    BZBJ0013+1910 & J001356.37+191041.9 & N/A & 3.48483333333 & 19.1783333333 & 0.477 & 42.0567302367 \\
    BZBJ0017+1451 & J001736.90+145101.7 & N/A & 4.40375 & 14.8505277778 & 0.303 & 41.2671294407 \\
    BZBJ0021-0900 & J002142.25-090044.4 & N/A & 5.42604166667 & -9.01230555556 & 0.648 & 41.8362628391 \\
    BZBJ0032-2849 & J003233.08-284920.4 & 3FGL J0032.3-2852 & 8.13779166667 & -28.8223055556 & 0.324 & 41.7686091928 \\
    BZBJ0040-2719 & J004016.42-271911.7 & N/A & 10.068375 & -27.3198888889 & 0.172 & 41.1872720594 \\
    BZBJ0047+3948 & J004755.22+394857.5 & 3FGL J0048.0+3950 & 11.9800833333 & 39.816 & 0.252 & 41.3333179422 \\    
    \hline
\end{tabular}
\label{tab:complete}
\end{center}
\end{table*}

\section{Statistical Investigation} 
\label{sec:statistical_investigation}

In this section we characterize blazar spectral energy distributions 
using a variety of statistical analyses to determine the most significant trends.
The idea is to base this characterization on the data rather than
fitting arbitrary functional forms to the SEDs.

\subsection{Clustering Analysis} 
\label{sub:data_clustering}
\begin{figure}[hbpt]
\begin{center}
    \includegraphics[width=0.99\columnwidth]{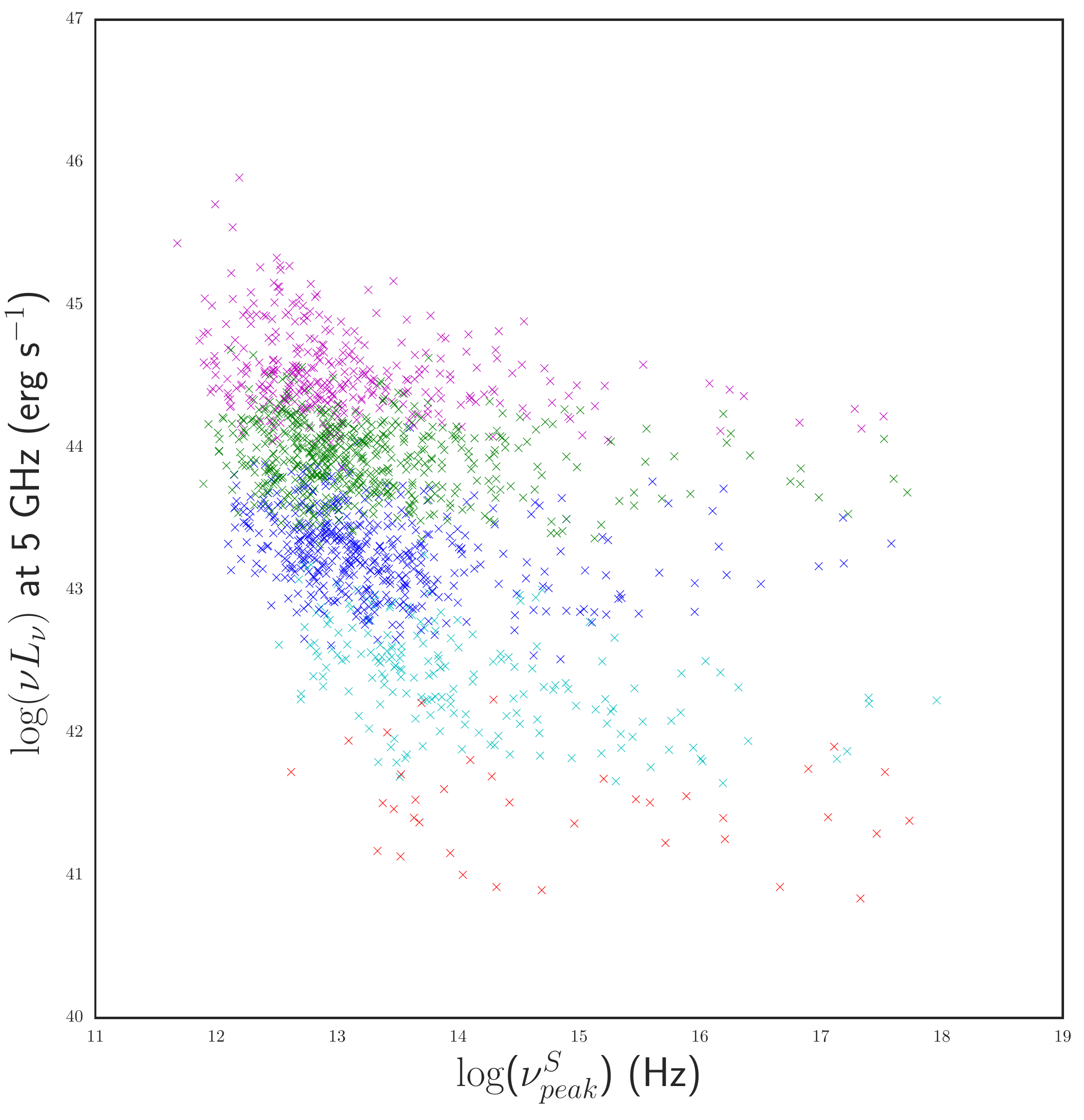}
    \caption{
    Radio luminosity versus synchrotron peak frequency for the $1590$ 
    blazars with radio data and complete optical or complete infrared data, 
    with colors defined by the radio luminosity bins 
    in Figure \ref{fig:radio_clustering}. The synchrotron
    peaks were determined via a second order polynomial fit 
    to the radio-through-optical SED, 
    and are probably uncertain by at least one decade 
    (obtained by refitting and measuring the shift in peaks after removing WISE 
    or SDSS data from the 54 complete SEDs with well-defined peaks).
    Even with large scatter, the synchrotron peak frequency rises smoothly 
    as the radio luminosity decreases (coefficient of correlation $\rho=-0.476$).
    }
\label{fig:radio_sync_peak}
\end{center}
\end{figure}

\begin{figure}[hbpt]
\begin{center}
    \includegraphics[width=0.99\columnwidth]{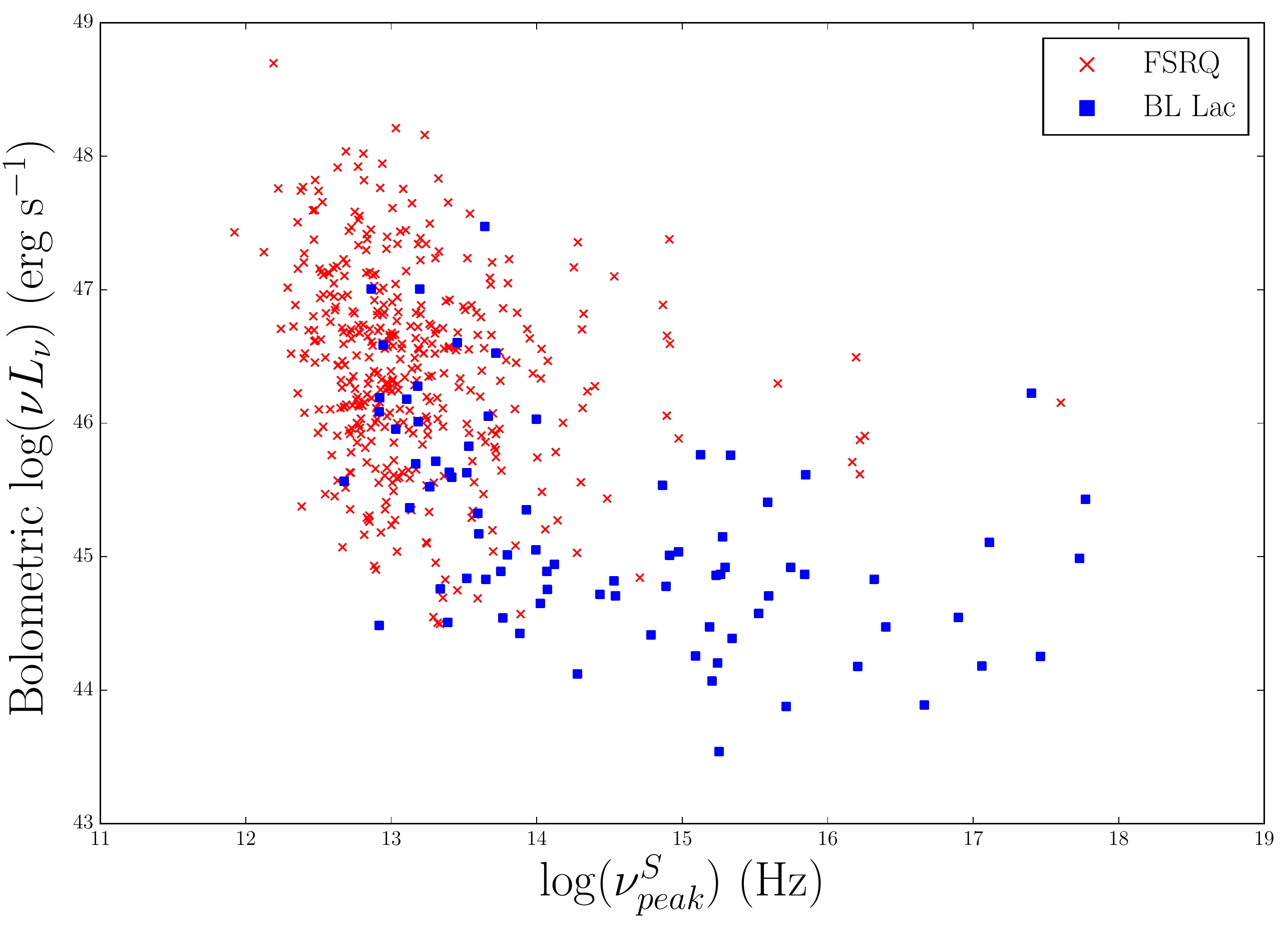}
     \caption{
    Bolometric luminosity versus synchrotron peak frequency for the $480$ blazars 
    with full radio, infrared and optical data. 
    $401$ red crosses represent flat-spectrum radio quasars 
    (i.e., blazars with broad emission lines) 
    and the $79$ blue filled squares are BL Lacs 
    (i.e., with no or weak emission lines). 
    Bolometric luminosity was determined from the peak $\nu L_{\nu}$ value;
    note that this is essentially the $\gamma$-ray luminosity for low-frequency-peaked,
    luminous blazars, or
    the X-ray luminosity for high-frequency-peaked BL Lac objects.
    As in Figure \ref{fig:radio_sync_peak}, synchrotron peak frequencies were found 
    via a second order polynomial fit to the radio-through-optical SED. 
    The same anti-correlation of synchrotron peak with luminosity is seen 
    ($\rho=-0.533$), 
    and there is no particular separation of FSRQs and BL Lacs, 
    although the former are more luminous on average than the latter. 
    }
\label{fig:bolometric_sync_peak}
\end{center}
\end{figure}

Previous works showed that radio luminosities of blazars are anti-correlated 
with the frequencies of their synchrotron peaks
\citep{sambruna96,fossati97,fossati98}. 
In our much larger sample, 
we see the same trend between radio luminosity and the synchrotron peak in the SED.
Specifically, Figure~\ref{fig:radio_sync_peak} illustrates
the anti-correlation between radio luminosity and the synchrotron peak; 
the large scatter in the latter is partly due to the difficulty in defining the peak frequency
from fewer than a dozen data points. 
The bolometric luminosity is similarly anti-correlated with synchrotron peak
(Figure~\ref{fig:bolometric_sync_peak}), with no sharp distinction between 
BL Lacs and FSRQs.

\begin{figure}[hbpt]
  \begin{center}
    \includegraphics[width=0.99\columnwidth]{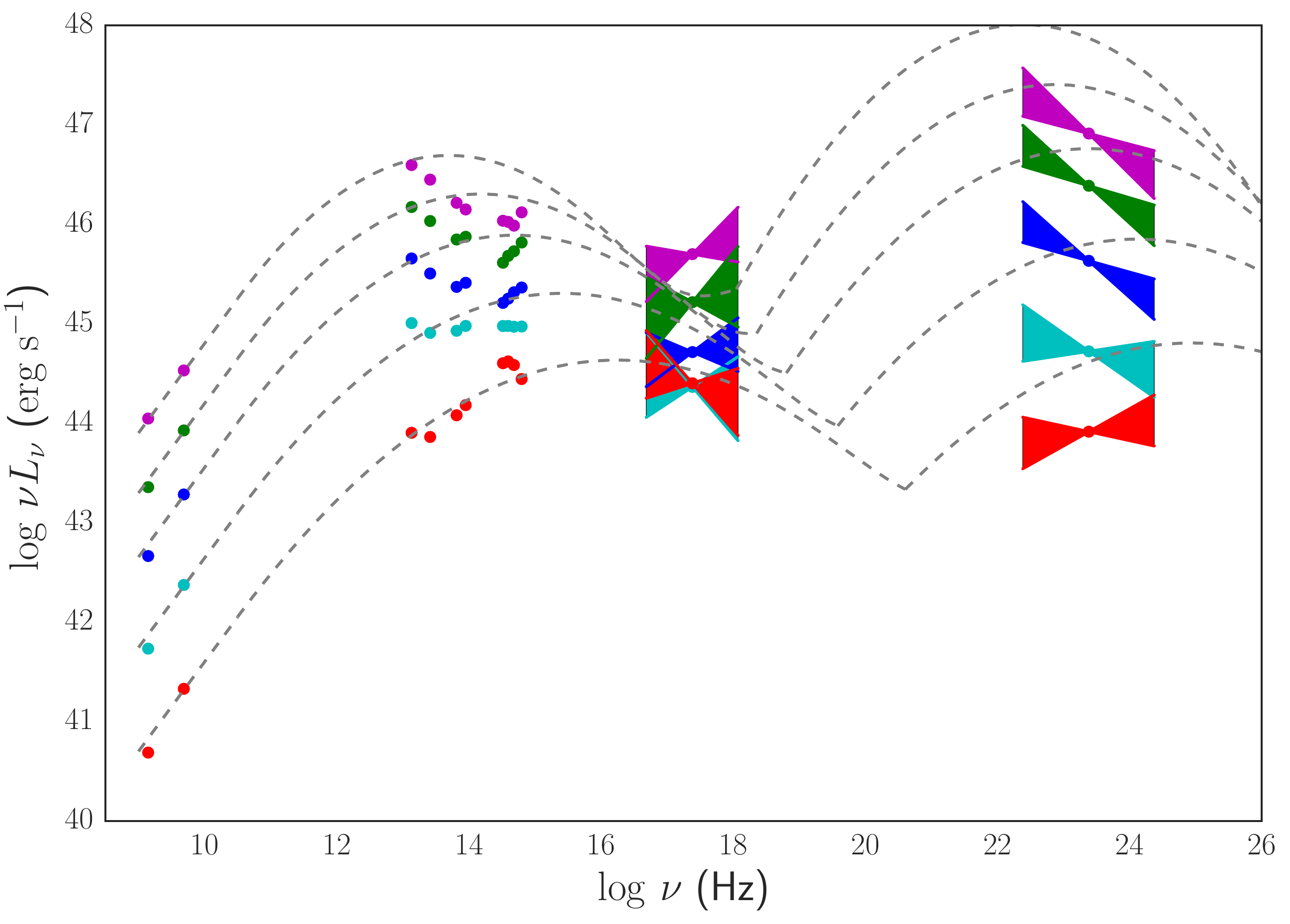}
  \end{center}
  \caption{
    Spectral energy distributions of $2214$ blazars sorted into five bins 
    of radio luminosity (purple, red, green, cyan, and blue, 
    in order of decreasing luminosity at 1.4~GHz);
    these were the natural clusters of K-nearest neighbors, 
    with $200$ to $700$ objects per luminosity bin of a decade or less in 
    luminosity.
    Median luminosity and X- and $\gamma$-ray slopes are plotted for each bin,
    with one standard deviation uncertainties. 
    The anti-correlation of synchrotron peak with luminosity is visible, 
    although a big blue bump (at high luminosities) 
    and host galaxy emission (in the lowest luminosity bin) 
    are comparable to the synchrotron contribution. 
    The analytic form SED proposed by \citet{fossati98} based on $126$ blazars
    ({\it dashed lines}) clearly does not fit the more extensive data.
}
\label{fig:radio_clustering}
\end{figure}

Figure~\ref{fig:radio_clustering} shows the SEDs of the 2214 blazars with radio, 
infrared and optical data, sorted into five bins of radio luminosity 
(roughly $200$ to $700$ objects per bin). 
The anti-correlation of synchrotron peak with luminosity is visible, 
even though a big blue bump from an accretion disk (at high luminosities) and 
host galaxy emission (in the lowest luminosity bin)
are comparable to the synchrotron contribution. 
The analytic form \citet{fossati98}fitted to $126$ blazar SEDs, 
only $34$ of which included $\gamma$-ray fluxes,
does not fit the more extensive data.  Neither are the data well fit by the updated parameters of \citet{donato01}, although the fit to the $\gamma$-ray data is slightly better.
$2160$ of \totnum\ blazars lack data in one or more bands; 
we show in Section~\ref{sub:missing}
that the blazars with missing data are not 
significantly
different
from the blazars that have those data, 
and so their absence does not affect the overall SED shape.

The SEDs for this large sample are not as well separated
as the well-known plots from \citet[Figure~12]{fossati98} may have made them 
appear; outside the radio band (which by definition has distinct bins), 
the SEDs overlap substantially, as indeed they did in the original Fossati sample.
In other words, the scatter at optical or $\gamma$-ray wavelengths is larger than 
the scatter within a given radio luminosity bin. 
Furthermore, the clustered SEDs differ significantly
from the analytic \citet{fossati98} curve.
 
\begin{figure}[hbpt]
  \begin{center}
    \includegraphics[width=0.99\columnwidth]{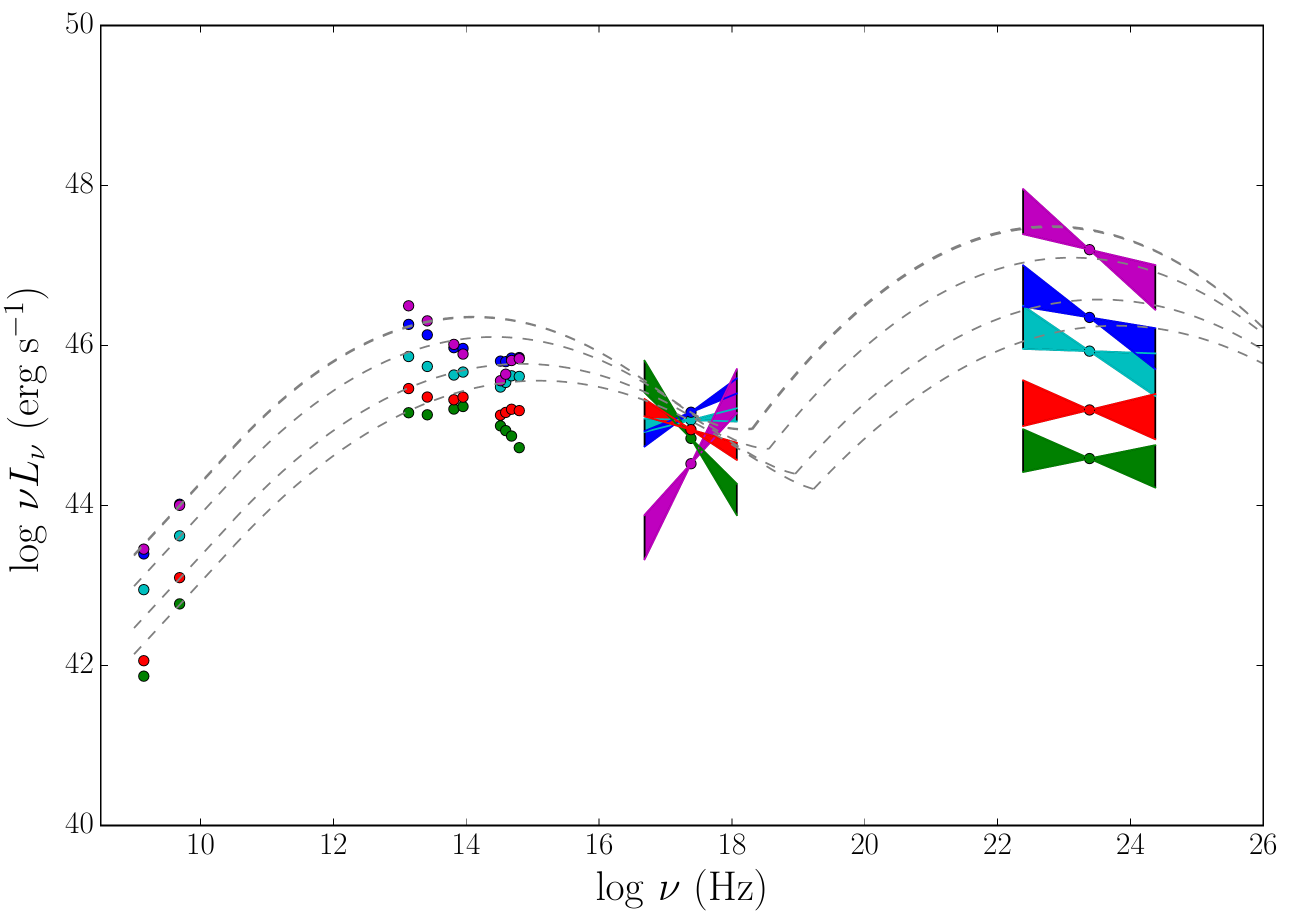}
  \end{center}
  \caption{
    Blazar SEDs binned according to X-ray spectral index, 
    showing the clear trend of hardening X-ray spectral index 
    with increasing radio luminosity. 
    The median luminosity in each bin and the median X- and $\gamma$-ray slopes 
    are plotted, with one standard deviation uncertainties.
    The analytic SED shape proposed by \citet{fossati98}, 
    keyed to the 5 GHz radio frequency bin ({\it dashed lines}), 
    does not fit the data.
}
\label{fig:xray_clustering}
\end{figure}

The blazar sample can also be sorted naturally --- and slightly differently --- 
by X-ray spectral index, as illustrated in Figure~\ref{fig:xray_clustering}.
The X-ray spectral index varies through a large range,
depending on whether the low-energy or high-energy SED component is dominant.
If the synchrotron radiation dominates at X-ray energies, 
the X-ray spectrum falls steeply with increasing frequency (decreasing wavelength);
if instead a Compton-scattered component dominates,
the X-ray flux rises with frequency. 
The different X-ray slopes are reminiscent of the \citet{fossati98} SED pattern 
(even though that was based on a radio sorting),
with high-luminosity, low-frequency-peaked blazars having 
hard X-ray spectra that are part of the Compton-scattered component,
and low-luminosity, high-frequency-peaked blazars having 
soft X-ray spectra that are the high-energy tail of the synchrotron component.

\begin{figure}[hbpt]
\begin{center}
    \includegraphics[width=1.09\columnwidth]{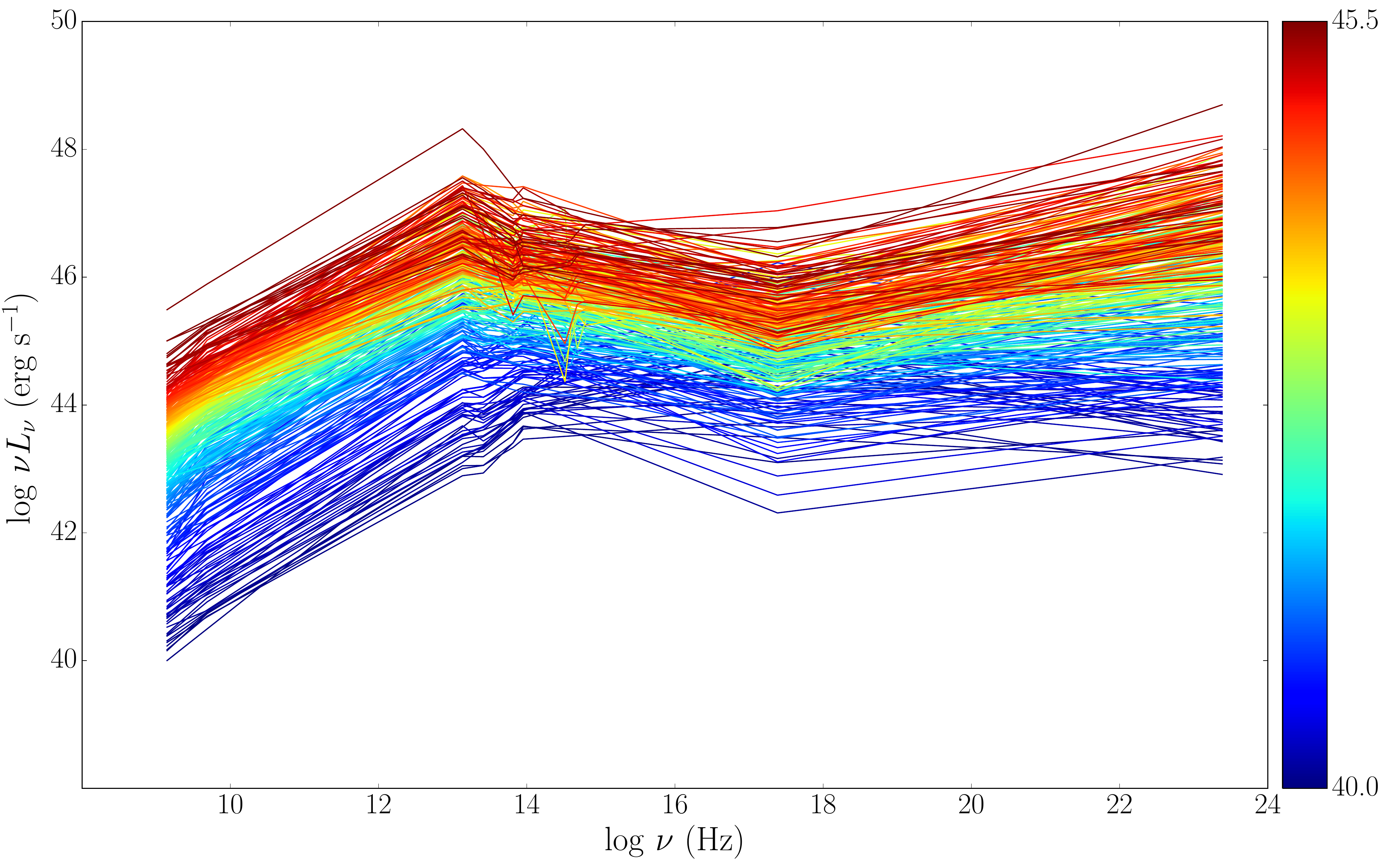}
     \caption{
    Individual SEDs of $347$ blazars that have 
    radio, X-ray and $\gamma$-ray data, and either WISE or SDSS data (or both). 
    Data points are connected using straight lines without binning in order to 
    avoid any assumption about the SED shape. 
    The color is scaled to the radio luminosity.
    The overall trend of decreasing synchrotron peak with increasing luminosity 
    is clear, as is the large dispersion around the average SED shape.
    }
\label{fig:SED_color_gradient_orig}
\end{center}
\end{figure}

The trend of SED shape with blazar luminosity is clearly seen, 
with the more $\gamma$-ray luminous objects having low-frequency synchrotron peaks 
and hard X-ray spectra, and the less luminous objects 
having high-frequency synchrotron peaks and soft X-ray spectra. 
A plot of the individual (unbinned) SEDs 
is shown in Figure~\ref{fig:SED_color_gradient_orig}.
These shapes are not easily described by an
analytic formula, even generalizing from \citet{fossati97},
nor is the scatter as small as that work suggested.
Accordingly, we looked for a general way to describe the SED shapes and
their trends with luminosity. 


\subsection{Multi-band correlations} 

\subsubsection{Principal Component Analysis} 
\label{sub:principal_component_analysis}
Principal Component Analysis (PCA) is a useful 
non-parametric method that can be applied to find correlations among fluxes at different wavelengths.
Briefly, PCA finds and orders the strongest correlations in an $n$-dimensional space 
(here $n=12$), and returns the basis vectors and their associated variances.
By definition, the components, or basis vectors, 
are orthogonal to one another and thus uncorrelated. 
The appropriate number of basis vectors is always less than or equal 
to the dimensionality of the original space.
In short, PCA redefines the axis of the parameter space so that 
the basis vectors best define the trends inherent in the data. 

We first performed PCA on the set of \completenum\ blazars 
with complete data across all 12 bands,
in order
to obtain a reference set of basis vectors to help guide subsequent analyses. 
Table~\ref{tab:pca_orig} lists the coefficients for each of the four dominant basis vectors,
as well as the associated standard deviations, 
fractional variance and cumulative variance.
\begin{table}[hbpt]
    \caption{Coefficients and Variances of Basis vectors for \completenum\ Complete Blazar SEDs}
\begin{center}
\begin{tabular}{ccccc}
\hline
Band & BV 1 & BV 2 & BV 3 &  BV 4  \\
\hline
1.4GHz & -0.36978 & 0.23622 & 0.4954 & -0.28891 \\
5GHz & -0.36674 & 0.20313 & 0.3554 & -0.14724 \\
W4 & -0.33667 & 0.23729 & -0.2406 & 0.33268 \\
W3 & -0.32203 & 0.18565 & -0.2515 & 0.30176 \\
W2 & -0.27201 & 0.06799 & -0.2697 & 0.22561 \\
W1 & -0.27406 & 0.02293 & -0.259 & 0.12514 \\
z & -0.22425 & -0.46464 & 0.2136 & 0.25268 \\
i & -0.22964 & -0.40565 & 0.1093 & 0.10032 \\
r & -0.23103 & -0.32775 & 0.0894 & 0.06374 \\
g & -0.22527 & -0.29663 & 0.1485 & 0.05732 \\
X-Ray & -0.19481 & -0.35295 & -0.3604 & -0.51569 \\
$\gamma$-Ray & -0.33585 & 0.1946 & -0.2496 & -0.48532 \\
\hline
Std Dev & 2.498 & 0.69322 & 0.45604 & 0.38181 \\
Prop. Var. \footnote{The variance each basis vector contributes to the total variance.} 
& 0.844 & 0.06501 & 0.02814 & 0.01972 \\
Cum Var. \footnote{The cumulative variance of all previous basis vectors.}
& 0.844 & 0.90901 & 0.93714 & 0.95687 \\
\hline
\end{tabular}
\end{center}
\label{tab:pca_orig}
\end{table}

The first basis vector, which depends approximately equally on all 12 fluxes,
carries 84\% of the total variance in the data.
The second basis vector depends differently on the optical and X-ray fluxes 
(change of sign) and explains another 6\% of the variance;
the third basis vector emphasizes the radio, X-ray and $\gamma$-ray data.
Together, the first four basis vectors explain 96\% of the total variance in the blazar SEDs.

We were concerned that the optical fluxes might include a contribution 
from host galaxy light (in low-luminosity blazars) or 
from a luminous UV-emitting accretion disk (in high-luminosity blazars) 
that are thermal component contaminations not related to the blazar SED shape we are investigating.
Accordingly, we repeated the PCA analysis excluding the SDSS data, 
and indeed, the first basis vector now explains a slightly higher fractional variance (89\%);
in other words, the SEDs are slightly more uniform without the 
(possibly unrelated) optical points.
The sense of the second basis vector is also different, 
with opposite dependence on radio luminosity and with much higher variance, 
suggesting that the radio band indeed carries more information 
about the nonthermal jet emission
than the other bands. 
The coefficients and significances of these basis vectors (i.e., without SDSS data) are 
listed in Table~\ref{tab:pca_orig_no_SDSS}.    
\begin{table}[hbpt]
    \caption{Coefficients and variances of each basis vector for 338 blazars with complete data apart from SDSS}
\begin{center}
\begin{tabular}{ccccc}
\hline
Band         & BV 1 & BV 2 & BV 3 &  BV 4  \\
\hline
1.4GHz & -0.38915 & 0.39474 & 0.41777 & 0.06506 \\
5GHz & -0.3929 & 0.35039 & 0.31293 & 0.08492 \\
W4 & -0.38571 & 0.01232 & -0.27929 & 0.13689 \\
W3 & -0.3668 & -0.02357 & -0.27423 & 0.18416 \\
W2 & -0.30624 & -0.09628 & -0.21904 & 0.26454 \\
W1 & -0.2984 & -0.12845 & -0.17257 & 0.24365 \\
X-Ray & -0.21285 & -0.65685 & 0.65246 & -0.00855 \\
$\gamma$-Ray & -0.42001 & -0.27236 & -0.24951 & -0.70395 \\
\hline
Std Dev & 2.5732 & 0.52681 & 0.45663 & 0.36484 \\
Prop. Var. \footnote{The variance each basis vector contributes to the total variance.}
& 0.8901 & 0.03731 & 0.02803 & 0.01789 \\
Cum Var. \footnote{The cumulative variance of all previous basis vectors.}
& 0.8901 & 0.92738 & 0.95541 & 0.9733 \\
\hline
\end{tabular}
\end{center}
\label{tab:pca_orig_no_SDSS}
\end{table}

It is also possible that the thermal dust torus emission 
or even the accretion disk contributes to the infra-red data. We subsequently
performed a PCA analysis without the WISE data, and found that 
the proportional variance of first basis vector drops to 80\%
from the original 84\%,
indicating a loss of information on correlation and not the removal 
of an independent component like a torus.
The coefficients and significances of these basis vectors are 
listed in Table~\ref{tab:pca_orig_no_WISE}. 
\citet{plotkin12} discussed in detail that the BL Lacs do not have thermal torus emission in WISE.

\begin{table}[hbpt]
    \caption{Coefficients and variances of each basis vector for  blazars with complete data apart from WISE}
\begin{center}
\begin{tabular}{ccccc}
\hline
Band         & BV 1 & BV 2 & BV 3 &  BV 4  \\
\hline
1.4GHz & -0.46281 & 0.3921 & -0.22 & 0.24046 \\
5GHz & -0.45512 & 0.3287 & -0.1758 & 0.20664 \\
z & -0.29242 & -0.404 & -0.2508 & -0.56169 \\
i & -0.29397 & -0.3677 & -0.1297 & -0.1048 \\
r & -0.2922 & -0.2975 & -0.1239 & 0.15678 \\
g & -0.28642 & -0.2529 & -0.1662 & 0.12781 \\
X-Ray & -0.24781 & -0.336 & 0.499 & 0.48837 \\
$\gamma$-Ray & -0.41324 & 0.2712 & 0.6718 & -0.47021 \\
\hline
Std Dev & 1.9729 & 0.65106 & 0.4172 & 0.36496 \\
Prop. Var. \footnote{The variance each basis vector contributes to the total variance.}
& 0.8073 & 0.08792 & 0.0361 & 0.02763 \\
Cum Var. \footnote{The cumulative variance of all previous basis vectors.}
& 0.8073 & 0.89522 & 0.9313 & 0.95894 \\
\hline
\end{tabular}
\end{center}
\label{tab:pca_orig_no_WISE}
\end{table}

\subsubsection{Assessing the effect of missing data on estimates of SED shape}
\label{sub:missing}

We want to use existing data to study blazar SEDs and to form 
an estimator for SED shape using the observed correlations. 
However, with data that are basically flux-limited,
the absence of data in a given band might not be uncorrelated with 
that object's SED shape.

In particular, many of the blazars in our sample lack data in one or more 
of the 12 bands, with only \completenum\ having data in every band.
Instead of restricting the sample to that handful of objects, 
it makes sense to use all the available information but only after
ascertaining that the partial data do not bias the sample.

\begin{figure*}[hbpt]
  \begin{center}
    \includegraphics[width=2.09\columnwidth]{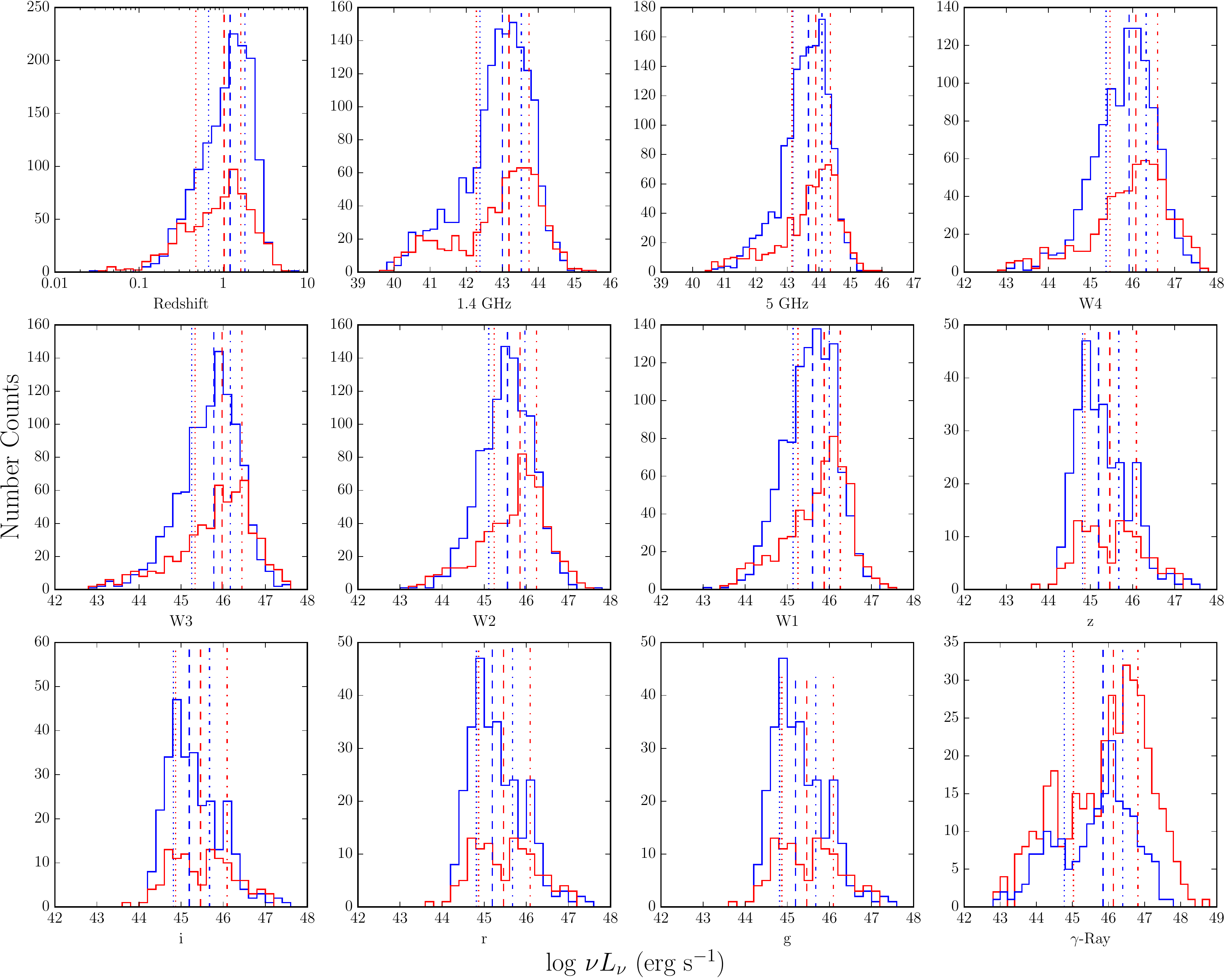}
  \end{center}
  \caption{Histograms comparing the multiwavelength flux distributions of blazars with ({\it red}) 
  and without ({\it blue}) 
  X-ray 
  data. The vertical lines indicate the mean fluxes ({\it dashed lines}) and the 
          25th ({\it dotted lines}) and 75th percentiles ({\it dot-dash lines}). 
}
\label{fig:hist_missing}
\end{figure*}

We first plot several histograms comparing objects with and without data in a specific band.
In Fig.~\ref{fig:hist_missing} we show the particular case of blazars with and without X-ray 
data. As is
evident from the dotted, dashed and dot-dash lines in the plot (corresponding to the
25th, 50th and 75th percentiles), the two subgroups have visually different distributions.
A series of Kolmogorov-Smirnov (KS) tests were carried out,
and test statistics are given in Table~\ref{tab:ks}.
\begin{table*}[hbpt]
    \caption{Kolmogorov-Smirnov Probability of values drawn from the same distribution}
\begin{center}
\begin{tabular}{cccccccccccccc}  
\hline
Band \footnote{The rows are bands omitted, and the columns are bands in consideration}
& Redshift &1.4~GHz & 5~GHz & W4 & W3 & W2 & W1 & z & i & r & g & X-ray & $\gamma$-ray \\
\hline
WISE &  1.3e-17 & 1.4e-18 & 0.0e+00 & NA & NA & NA & NA & 1.3e-01 & 1.5e-01 & 1.2e-01 & 4.0e-07 & 5.3e-17 & 2.7e-07 \\
SDSS &  1.5e-02 & 1.5e-01 & 6.5e-01 & 4.8e-01 & 3.5e-01 & 1.2e-01 & 5.9e-293 & NA & NA & NA & NA & 3.6e-01 & 4.2e-01 \\
X-ray &  4.2e-06 & 2.7e-05 & 3.0e-11 & 1.7e-06 & 1.8e-06 & 1.6e-06 & 1.8e-06 & 1.2e-01 & 1.8e-01 & 6.6e-02 & 6.2e-02 & NA & 5.8e-73 \\
$\gamma$-ray &  1.9e-02 & 8.5e-03 & 5.6e-18 & 1.1e-17 & 1.8e-17 & 2.2e-17 & 1.5e-01 & 4.3e-01 & 3.5e-01 & 5.6e-01 & 2.7e-83 & 0.0e+00 & NA \\
\hline
\end{tabular}
\end{center}
\label{tab:ks}
\end{table*}

While every band (other than radio at 1.4~GHz) has some missing data, 
missing optical data are in fact the worst problem,
as only $432$ blazars have complete SDSS 
photometry\footnote{The number is roughly as expected given the SDSS footprint 
compared to the area of radio- and $\gamma$-ray-surveyed sky.}
(the fewest of all wavelengths). 
By comparing the coefficients and the variances of the 
PCA basis vectors
in Tables~\ref{tab:pca_orig} and \ref{tab:pca_orig_no_SDSS},
it is clear that the intrinsic correlations among
the SED bands did not change much after the SDSS data were removed.
In other words, the $338$ blazars with complete data excepting the optical have
SEDs that are essentially indistinguishable from the \completenum\ full-data blazars.

We then repeated the PCA analysis to assess the effect of missing $\gamma$-ray data,
since only $571$ blazars (measured redshifts) have Fermi data.
For the $98$ blazars with complete data excepting the $\gamma$-ray, 
the basis vectors again had similar coefficients and variances, thus indicating 
they are indistinguishable from the \completenum\ full-data blazars.

Table \ref{tab:pca_summary} shows the coefficients for the first basis vector 
for blazars that have
complete data minus one or more bands. 
The significance of each basis vector and the overall PCA result
does not change much if one or more bands are excluded. 
This indicates that the overall SED shapes with and without 
certain bands do not look very different, 
i.e., the blazars with missing data do not strongly bias the results.

\begin{table*}[hbpt]
    \caption{Values of Basis vector 1 for Samples Missing Data in Various Bands}
\begin{center}
\begin{tabular}{cccccccccc}\\
\hline
Omitted bands & None & SDSS & $\gamma$-ray & X-ray & SDSS \& $\gamma$-ray & SDSS \& X-ray & $\gamma$-ray \& X-ray & SDSS, $\gamma$-ray \& X-ray \\
\hline
\# blazars & \completenum\ & 338 & 98 & 83 & 561 & 492 & 293 & 1603 \\
\hline
1.4~GHz & -0.36978 & -0.38915 & -0.39152 & -0.36792 & 0.43105 & -0.39492 & -0.43995 & -0.3944 \\
5~GHz & -0.36674 & -0.39290 & -0.38858 & -0.36582 & 0.43476 & -0.39925 & -0.44396 & -0.3912 \\
W4 & -0.33667 & -0.38571 & -0.35585 & -0.33694 & 0.42352 & -0.39699 & -0.43835 & -0.3597 \\
W3 & -0.32203 & -0.36680 & -0.34062 & -0.32370 & 0.40302 & -0.37845 & -0.41804 & -0.3456 \\
W2 & -0.27201 & -0.30624 & -0.28843 & -0.27459 & 0.33706 & -0.31661 & -0.35014 & -0.2938 \\
W1 & -0.27406 & -0.29840 & -0.29007 & -0.27791 & 0.32809 & -0.30996 & -0.34242 & -0.2968 \\
z & -0.22425 & NA & -0.24102 & -0.23598 & NA & NA & NA & -0.2537 \\
i & -0.22964 & NA & -0.24565 & -0.24235 & NA & NA & NA & -0.2596 \\
r & -0.23103 & NA & -0.24708 & -0.24458 & NA & NA & NA & -0.2623 \\
g & -0.22527 & NA & -0.24128 & -0.24051 & NA & NA & NA & -0.2584 \\
X-ray & -0.19481 & -0.21285 & -0.20706 & NA & 0.23285 & NA & NA & NA \\
$\gamma$-ray & -0.33585 & -0.42001 & NA & -0.35645 & NA & -0.43074 & NA & NA \\
\hline
Prop. Var. \footnote{The fractional variance of the first basis vector.}
& 0.844 & 0.8901 & 0.8452 & 0.8419 & 0.890 & 0.9255 & 0.9342 & 0.8435 \\
\hline
\end{tabular}
\end{center}
\label{tab:pca_summary}
\end{table*}

\subsection{Imputation}

Since PCA works only with complete data (by the nature of the algorithm),
we considered measures to fill in the missing data. 
Two methods, namely, Multiple Imputation and Sequential K-Nearest Neighbors, 
were tried. 

Multiple Imputation is a model-based approach to 
estimating the potential value of a missing data point. 
A series of iterative regression analyses were run 
in which one particular band was used as the dependent variable 
and the other bands were used as independent variables. 
The regression model was generated with all available data,
and if the dependent band had any missing values, 
these values were estimated by the regression using data from all the other bands. 

\begin{figure}[hbpt]
  \begin{center}
    \includegraphics[width=1.05\columnwidth]{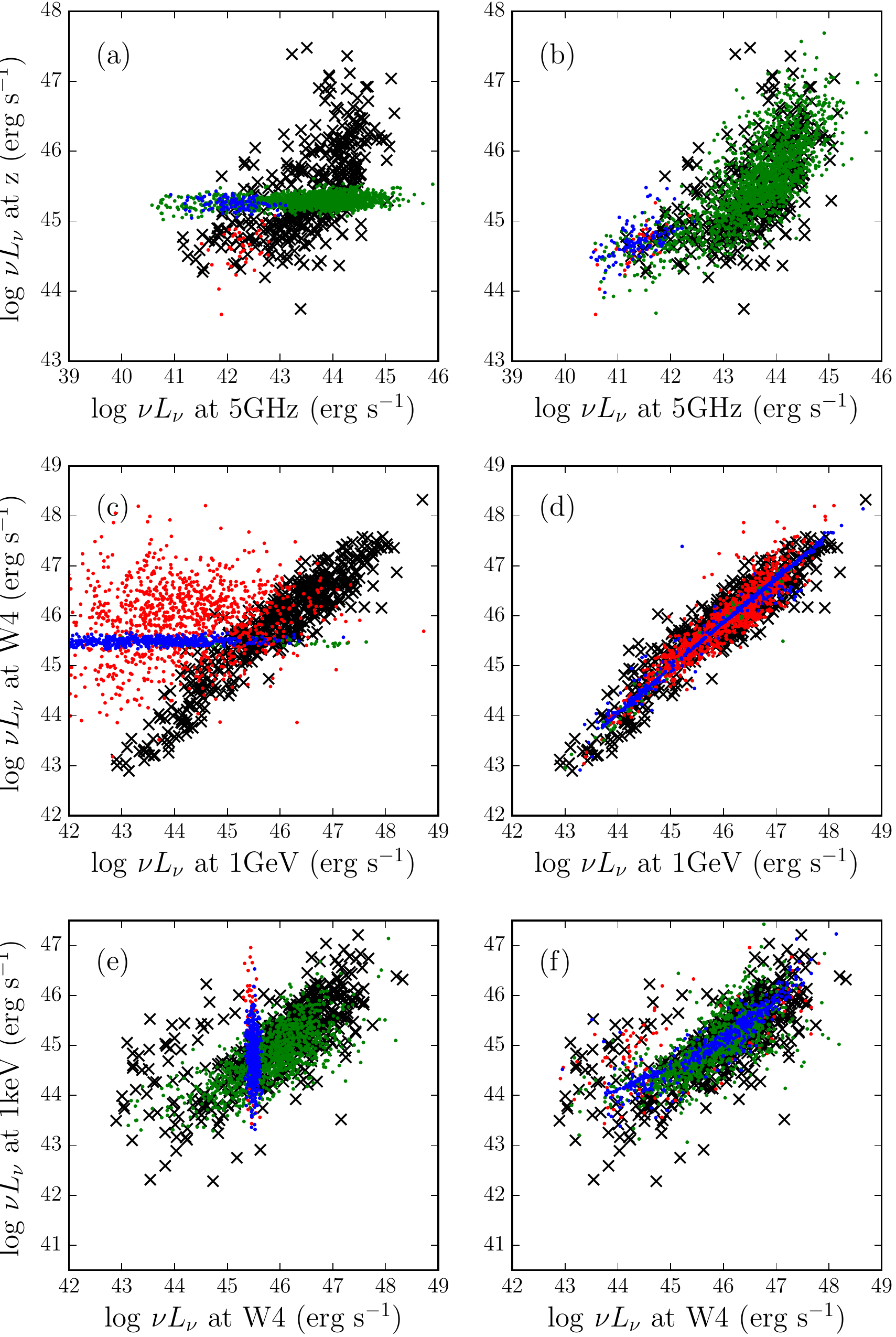}
  \end{center}
  \caption{ Some of the correlation plots of Multiple Imputation ({\it left}) 
            and K-nearest Neighbors 
            ({\it right}) using all the data available (up to 12 bands). 
            In these plots, the black crosses are data complete in both bands, 
            red dots are imputed data originally missing in the band on the $x$-axis; 
            green dots are imputed data originally missing in the $y$-axis. 
            Blue dots are imputed data missing in both plotted bands. 
            The Multiple Imputation approach
            creates trends not present in the original data (horizontal and 
            vertical strips), while the KNN-imputed values follow the existing trends.
  }
\label{fig:mi_knn}
\end{figure}

We saw unphysical results from the Multiple Imputation approach, 
where many of the correlation plots showed 
strange vertical and/or horizontal strips
(see Figure~\ref{fig:mi_knn}, panels $a$, $c$, $e$).

These linear features look very different from the original data,
due to the fact that Multiple Imputation
uses data from all 12 bands of all \totnum\ blazars
to over-specify the structure to the point that 
the model does not have enough freedom 
to produce useful predictions; 
instead, it produces repetitive predicted 
fluxes in line with the populated information. 

Sequential K-Nearest Neighbors (KNN), on the other hand, 
imputes a missing data point from the closest $k$ 
data in an $(N - 1)$-dimensional space defined by the vectors not being considered. 
In our analysis, we used the generally assumed number value of $k=10$ \citep{feigelson12}.
Other $k$ values did not produce better results.
The imputed value of the missing point is equal to the 
inverse distance-weighted mean of its its neighbor's values, 
and the distance measures considered here are ``statistical'' distances, 
i.e., the mean and the co-variance.

\begin{figure*}[hbpt]
  \begin{center}
    \includegraphics[width=2.1\columnwidth]{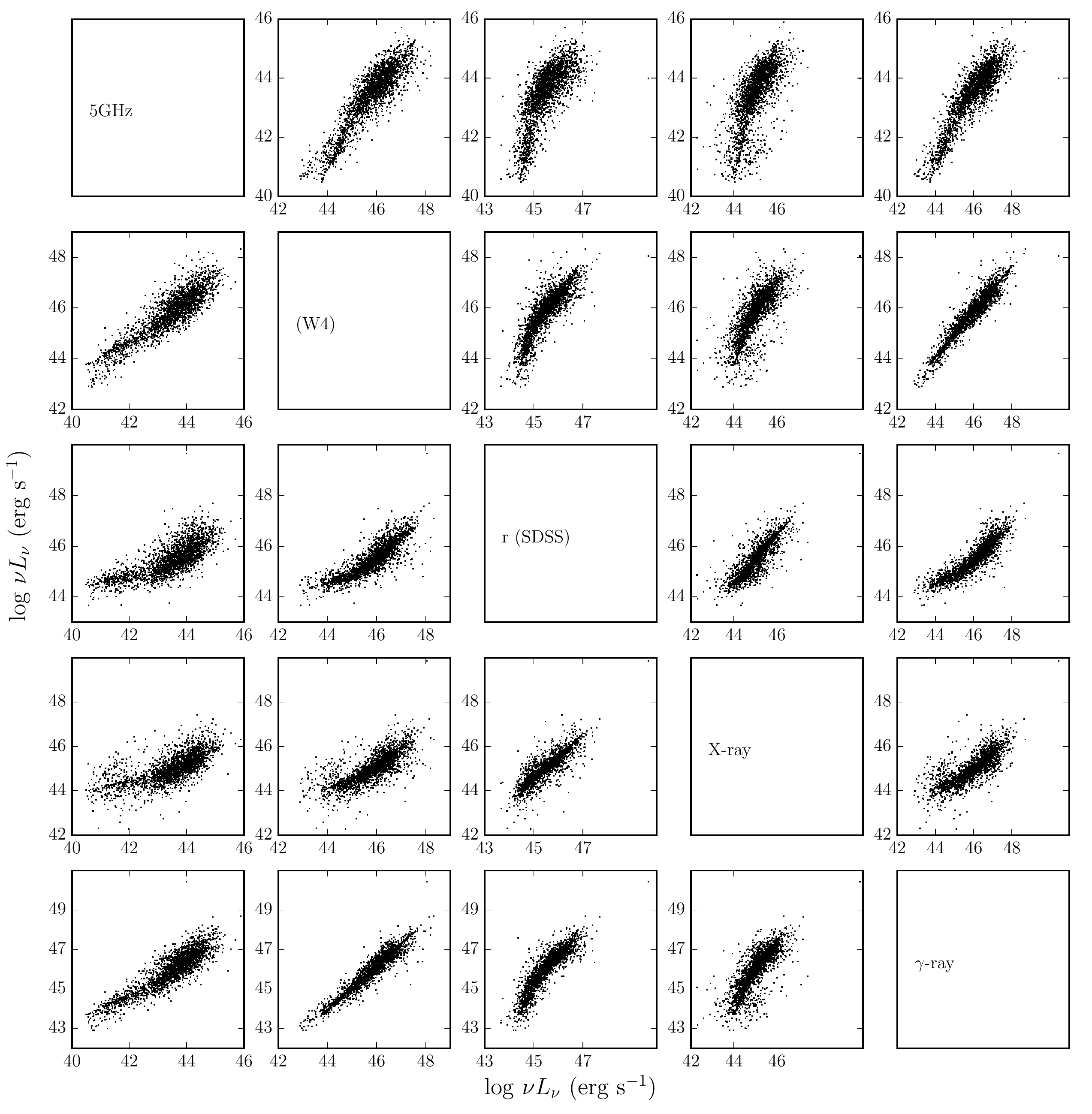}
  \end{center}
  \caption{KNN pairs plots with imputed data. 
  The general linear trends are conserved in all cases.}
\label{fig:knn}
\end{figure*}

Sequential KNN gave much better results than Multiple Imputation, as
the imputed data very nicely follow 
the form of the originally populated values as we 
would have 
expected 
(Figures~\ref{fig:mi_knn} and \ref{fig:knn}, panels $b$, $d$, $f$).

One potential drawback of KNN is that 
it mimics our data in \emph{every} respect,
reducing the variance artificially.
We proceed using the KNN method to impute missing data, 
and interpret the results with appropriate caution. 
Figure~\ref{fig:knn} shows the 2-band correlations with real and imputed data.

We next ran the PCA analysis on the real plus imputed data.
As expected, the variance explained by the first basis vector is larger 
because KNN-imputed data mimic the original data.
Coefficients and significances of the newly generated basis vectors 
are shown in Table \ref{tab:pca_knn}.
\begin{table}[hbpt]
    \caption{Coefficients and variances of each basis vector for original plus KNN-imputed data}
\begin{center}
\begin{tabular}{ccccc}
\hline
Band & BV 1 & BV 2 & BV 3 &  BV 4  \\
\hline
1.4GHz & -0.4075 & 0.549 & 0.3379 & 0.0221 \\
5GHz & -0.40487 & 0.4616 & 0.1664 & 0.00598 \\
W4 & -0.37366 & -0.1263 & -0.3949 & 0.37292 \\
W3 & -0.36218 & -0.1712 & -0.2914 & 0.10959 \\
W2 & -0.30287 & -0.2446 & -0.0519 & -0.30574 \\
W1 & -0.27747 & -0.2725 & 0.0128 & -0.55548 \\
X-Ray & -0.22374 & -0.4419 & 0.6861 & 0.06391 \\
$\gamma$-Ray & -0.41438 & -0.1836 & -0.1529 & 0.14074 \\
\hline
Std Dev & 2.3654 & 0.58008 & 0.38865 & 0.287 \\
Prop. Var. & 0.8834 & 0.05313 & 0.02385 & 0.013 \\
Cum Var. & 0.8834 & 0.93654 & 0.96039 & 0.9734 \\

\hline
\end{tabular}
\end{center}
\label{tab:pca_knn}
\end{table}

This PCA result shows the same correlations as for 
the original \completenum-complete blazars. 
Of course, it does rely heavily on those \completenum\ blazars,
but in fact this PCA uses available data from almost every single blazar, 
and the result is far more representative and less biased. And the fact that
the first basis vector still explains approximately 
88\% of the variation in our original data
tells us that all the bands load more or less evenly on this basis vector. 

The remaining basis vectors give us a sense of the natural band grouping 
and indicate how much explanatory power these groups provide on their own. 
Since the second basis vector is dominated by the radio bands,
considering the radio in its own right is an important step. 

\begin{figure*}[hbpt]
\begin{center}
    \includegraphics[width=1.99\columnwidth]{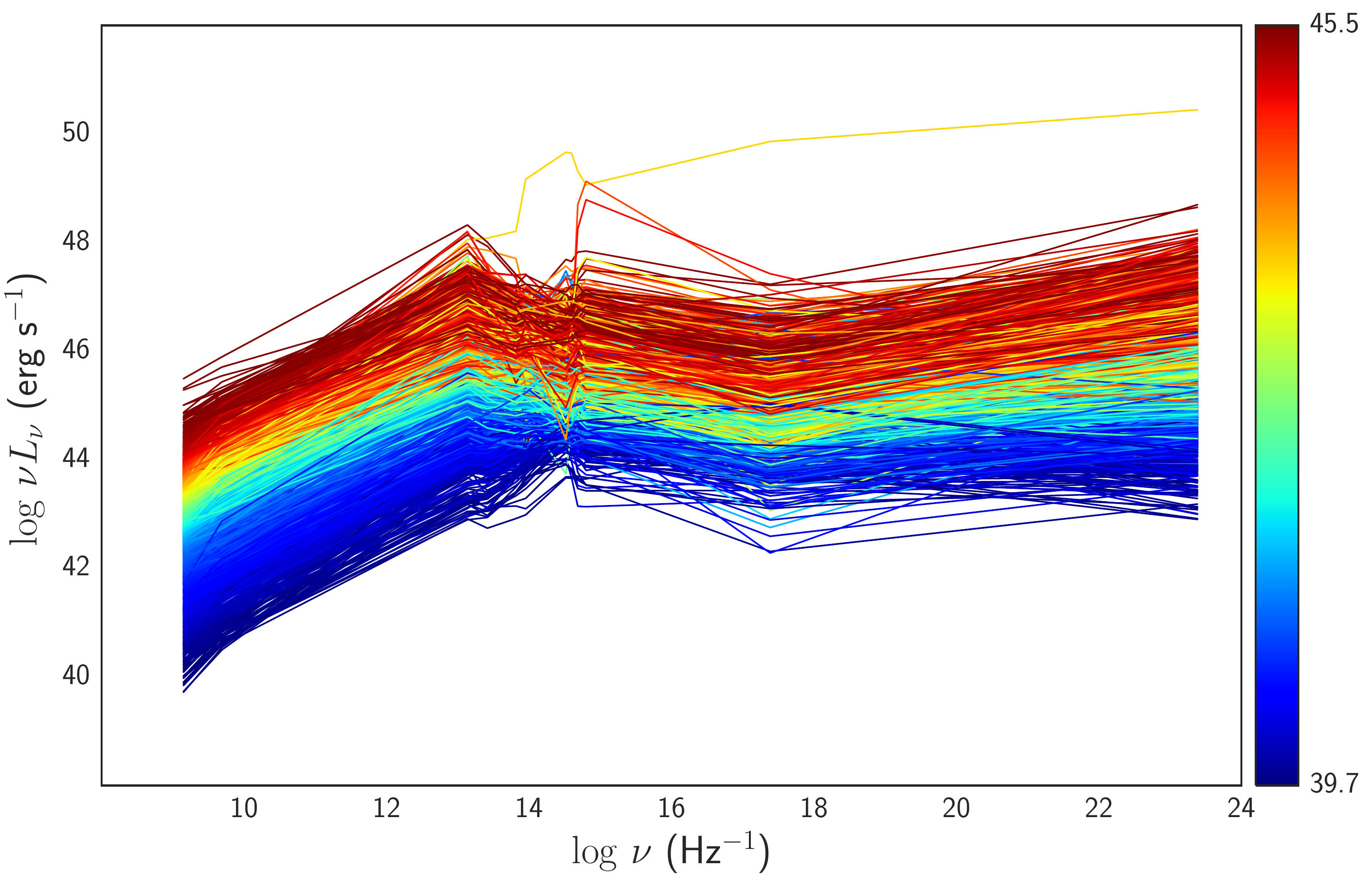}
    \caption{
    SEDs of the full sample of \totnum\ blazars, similar to Fig.~\ref{fig:SED_color_gradient_orig} 
    but including KNN-imputed data. The least-complete data 
    (blazars with only radio flux) are plotted first
    then blazars with increasingly complete SEDs, 
    such that the most-complete data (blazars with fluxes in 
    every band) are plotted last (i.e., on top of the other curves).
    Data points are connected with straight lines among the 12 bands. 
    There is much scatter but the synchrotron peak still decreases in frequency,
    and the $\gamma$-ray dominance increases, with increasing luminosity.
    }
\label{fig:SED_color_gradient}
\end{center}
\end{figure*}

Figure~\ref{fig:SED_color_gradient} shows the same array of SEDs 
as in Figure~\ref{fig:SED_color_gradient_orig} but including the KNN-imputed data. 
The anti-correlation between radio luminosity and 
synchrotron peak is preserved, as expected.

\section{Estimators} 
\label{sub:estimators}

Given the analysis presented in above,
we now describe an estimator for blazar luminosity at any wavelength based on 
the radio luminosity at 1.4~GHz.
Although we could use a different band as the input for an estimator 
(because the PCA analysis shows different wavebands to be highly correlated), 
using 1.4~GHz maximizes our completeness. 

The resultant functional form, obtained using the standard theory of linear models, 
is a first order polynomial in radio luminosity ($\log(\nu L_{\nu})$ at 1.4~GHz) 
and redshift ($\log(1+z)$): 
\begin{equation*}
    \log(\nu L_{\nu}) = c_0 + c_1 \log(1+z) + c_2 \log(\nu L_{\nu_{1.4GHz}}) .
\end{equation*}
The complete coefficients are listed in Table \ref{tab:estimator}.
\begin{table}[hpbt]
    \caption{Coefficients for the estimator}
\begin{center}
\begin{tabular}{ccccc}
\hline
Band & $c_0$  & $c_1$ & $c_2$ & $\sigma$ 
\footnote{Standard deviations in the Gaussian fit to the residuals of the estimator.} \\ 
\hline
5GHz & 5.692 & 0.614 & 0.876 & 0.173 \\
W4 & 27.977 & 3.085 & 0.393 & 0.39 \\
W3 & 28.593 & 3.03 & 0.376 & 0.355 \\
W2 & 30.955 & 2.464 & 0.322 & 0.366 \\
W1 & 30.358 & 1.874 & 0.341 & 0.422 \\
i & 39.29 & 2.822 & 0.121 & 0.455 \\
r & 38.909 & 2.838 & 0.131 & 0.402 \\
g & 38.967 & 3.199 & 0.127 & 0.434 \\
z & 37.824 & 3.311 & 0.153 & 0.476 \\
X-Ray & 37.762 & 2.458 & 0.152 & 0.518 \\
$\gamma$-Ray & 22.961 & 3.934 & 0.508 & 0.42 \\
\hline
\end{tabular}
\label{tab:estimator}
\end{center}
\end{table}

\begin{figure*}[hbpt]
\begin{center}
    \includegraphics[width=1.9\columnwidth]{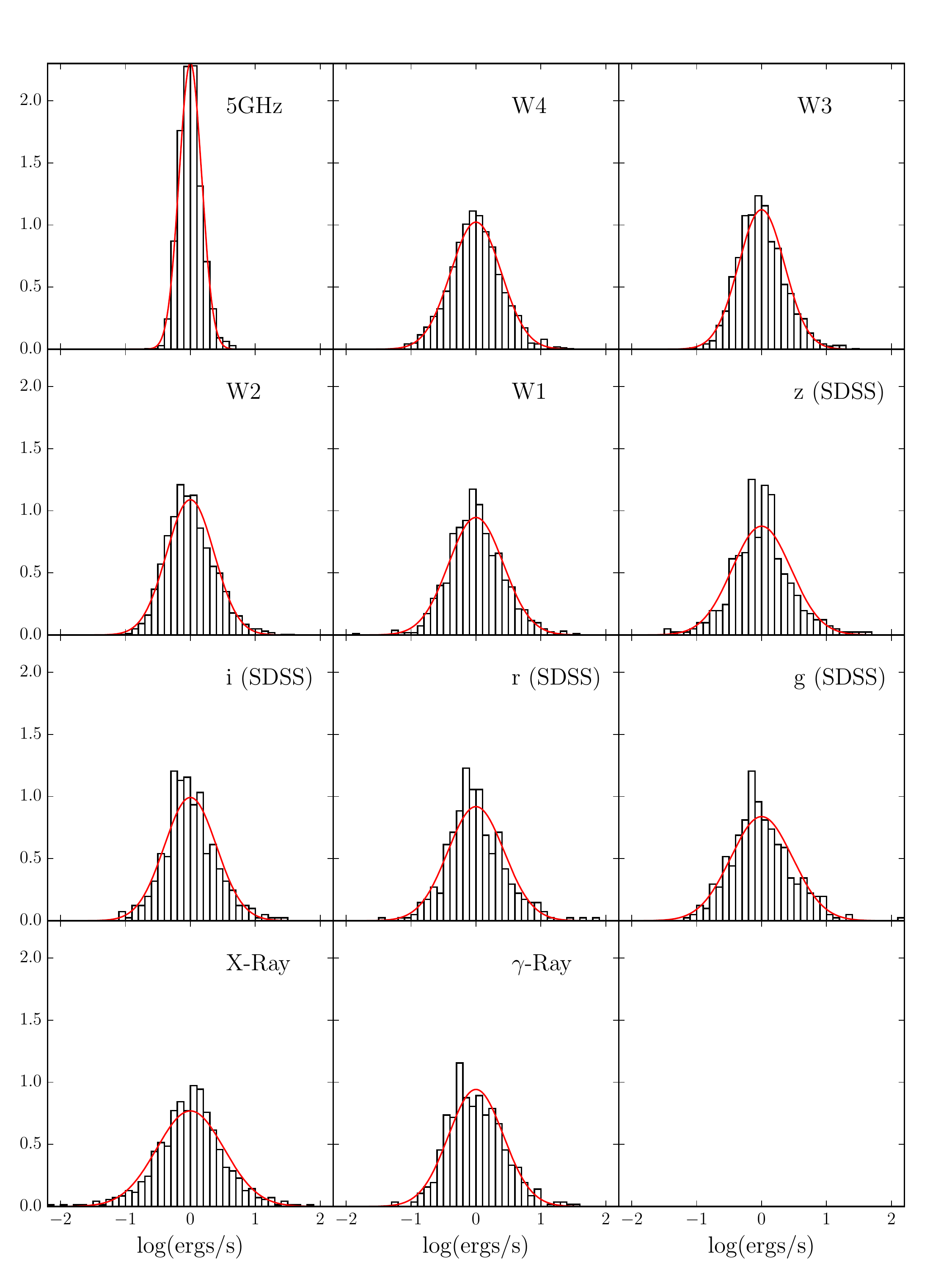}
    \caption{
    Residuals of the estimator with coefficients given in Table~\ref{tab:estimator},
    using redshift and 
    1.4-GHz 
    radio luminosity as inputs. 
    These are distributed around zero, with Gaussian half-widths of 1 sigma, 
    where 
    the $x$-axis is the difference 
    between the predicted value and the data value (difference in $\log{(\nu L_{\nu})}$);
    the total area is normalized to unity.}
\label{fig:estimator}
\end{center}
\end{figure*}

Figure~\ref{fig:estimator}
shows the residuals after applying the estimator to all blazars 
for which data are available in a particular band. 
The predicted estimator values were subtracted from the measured values 
to produce the residuals. 
The results are distributed in a well-behaved Gaussian around a mean value of zero, 
showing that the estimator is correctly predicting the distributions of blazar luminosities. 

A discussion of the origin of the correlations, which serves as the 
theoretical underpinning of this estimator, is in Section \ref{sec:conclusions}.
Better estimation with smaller residuals is
of course possible by increasing the degrees of freedom by adding more input terms, 
but our goal is not to fit perfectly each SED but to have a robust estimator that 
captures the general SED trends for the full population of blazars. 
With this estimator, the luminosity of any blazar with known radio luminosity 
and redshift can be predicted across $\sim$15 decades of the electromagnetic spectrum.
Such an estimator could be used, for example, to model the blazar population as a whole.

Finally, to test the efficiency/reliability of the statistical estimator (previously defined)
we performed the following test.
We considered the radio luminosities of the sources listed in our original blazar sample, 
and computed the expected values of the fluxes (K-corrected) 
at different frequencies and compared them with the observed values 
within a 3$\sigma$ interval. 
The reliability derived from this cross-validation test are reported 
in Table \ref{tab:reliability} for each energy range, respectively.

The reliability represents the fraction of sources for which the estimator correctly 
predicts the flux within 3$\sigma$ in a particular band. 
We note that blazars tend to be highly variable
objects, especially in the higher energy frequency bands,
and in addition, our large sample undoubtedly includes intrinsic scatter that 
was not as visible in the smaller samples studied previously. 
The discrepancy between the predicted and actual luminosities reflects both variational
and real difference. However, our goal is to correct distribution for the overall 
blazar population, not to recover accurate SEDs for individual objects, 
and the estimator, which is built on the largest and most complete 
collection of blazars to date, successfully achieves this goal. 

\begin{table}[hpbt]
    \caption{Reliability of the Estimator in respective energy bands}
\begin{center}
\begin{tabular}{cc}
\hline
Band & Reliability
\footnote{The reliability represents the fraction of sources for which 
the estimator correctly predicts the flux within 3~$\sigma$ in a particular band} \\ 
\hline
W4              &  68.8\% \\
W3              &  58.5\% \\
W2              &  16.9\% \\
W1              &  7.92\% \\
z               &  32.2\% \\
i               &  15.3\% \\
r               &  3.64\% \\
g               &  2.60\% \\
X-ray\footnote{An interval of 1~$\sigma$ is used for X-ray since the errors in count rates
are relatively big compared with the count rates themselves and using 3~$\sigma$ would 
result in unphysical answers for many objects.}
                &  9.80\% \\
$\gamma$-ray    &  14.0\% \\
\hline
\end{tabular}
\label{tab:reliability}
\end{center}
\end{table}


\section{Discussions and Conclusions} 
\label{sec:conclusions}

In this paper, we have characterized SEDs of \totnum\ blazars with complete redshift information,
using data from Roma-BZCAT, the largest 
and the most complete collection of known blazars.
Compared with previous work 
on blazar demographics
\citep{giommi94,stocke85,maraschi86,fichtel94, nieppola06, chen11, meyer11, ghisellini11, giommi12, giommi13},
this represents
the largest sample with more, and more uniform,
data from radio through $\gamma$-ray energies. 

Of the \totnum\ blazars in our sample, all have 
measured radio luminosities at 1.4~GHz, 
2056 have measured radio luminosities at 5~GHz, 
1628 have infrared fluxes
across all four WISE bands (W1, W2, W3, W4), 
432 have optical luminosities from SDSS (g, r, i, z), 
700 have X-ray luminosities at 1~keV, 
and 571 have $\gamma$-ray luminosites at 1~GeV.

Through non-parametric analysis of these
blazar SEDs, we have confirmed the connection between radio luminosity and SED shape, 
such that the synchrotron peak frequency decreases as 
the radio (also bolometric) luminosity increases.
Principal component analysis showed that
the first basis vector depends on the luminosities at various wavelengths, 
with roughly equal contributions to the variance;
this shows that 90\% of the correlation is explained by the bolometric luminosity.
In the second basis vector, the radio luminosities are the dominant contributors
to the variance (and have opposite sign to the other luminosities),
meaning it is reasonable to use the radio luminosity as the driving parameter.

We constructed a non-parametric estimator of the blazar SED based on
the radio luminosity (at 1.4~GHz) and the redshift, 
which are a key part of the first basis vector signal.
In a subsequent paper, this estimator will be combined with the 1.4-GHz radio luminosity function to study blazar 
demographics (Mao et al., in preparation). 

The SED trends we observe, called the ``blazar sequence''
by \citet{fossati98} and \citet{ghisellini98}, 
has been discussed extensively 
(see \citealp{ghisellini08, ghisellini09, ghisellini11, ghisellini12}, 
\citealp{giommi99, giommi12, giommi13}, \citealp{nieppola06, nieppola08} and \citealp{chen11}).
\citet{meyer11} extended the idea of the blazar sequence
to a broader ``blazar envelope'', which bounds two populations: 
highly aligned blazars with high synchrotron peaks (at all luminosities)
and misaligned radio galaxies with low synchrotron peaks and low luminosities.

The observed trends in blazar SED shape were
explained theoretically by \citet{ghisellini98} 
using a simple synchrotron inverse-Compton model.
Specifically, in luminous blazars with high photon energy densities 
due to accretion disk and line emission, 
the synchrotron-radiating electrons lose energy
through Compton up-scattering, producing strong $\gamma$-ray components
and lower synchrotron peak frequencies,
while the electron energy distribution remains hard in 
less luminous BL Lac objects, which lack strong line emission,
producing high-frequency synchrotron peaks and low $\gamma$-ray luminosities.
Thus the blazar sequence can be explained naturally. 

However, if high-luminosity, high-synchrotron-peak blazars are systematically 
missing from major surveys,
then the observed trend simply results from a selection effect.
\citealp{padovani07} proposed 
three tests that would prove the observed trend has a real physical basis:
(i) the existence of an anti-correlation between 
the synchrotron peak frequency and the bolometric observed luminosity; 
(ii) the absence of outliers from the correlation; 
(iii) and higher intrinsic numbers of high-frequency-peaked blazars 
relative to low-frequency-peaked ones.
Here we addressed the first test, showing through principal component analysis 
that the trend involves the overall luminosity (at all wavelengths sampled).
Furthermore, we checked that the correlation does not simply result from the 
common redshift dependence of luminosity at any wavelength
known as (the ``common-distance bias'', \citealp{pavlidou12}).
In particular, we permuted the redshifts
among the blazars and re-calculated luminosities 
at each wavelength; after intensive bootstrapping,
we found that the probability of measuring the correlation coefficient and slope 
we observe is less than 0.1\%.
We also tried assigning each blazar a random redshift from the observed distribution 
(instead of permuting the observed ones);
the probability of getting the observed correlation was again 
less than 0.1\%. Therefore, while redshift 
does introduce some correlation in luminosity space when none exists, 
the level of correlation seen in our sample 
is certainly driven by the redshift distribution,
thus being real under the assumption that our sample 
is representative of the whole blazar population.

The SED estimator we present forms the basis of a detailed Monte-Carlo simulation 
to probe the true volume densities of blazars,
i.e., to  answer precisely
whether high-frequency-peaked blazars are indeed more numerous than 
low-frequency-peaked blazars (Mao et al., in prep.) --- the third test mentioned above.
Another paper (Mao et al., in prep.) tests whether the blazars missing redshifts
are preferentially high-redshift, high-frequency-peaked BL Lac objects, 
i.e., the outliers identified in the second test mentioned above. 
So far, the first test --- of an intrinsic correlation between bolometric luminosity and SED shape ---
supports the intrinsic nature of the blazar sequence.

\section{Acknowledgments}

This publication makes use of data products from the Wide-field Infrared Survey Explorer, 
which is a joint project of the University of California, Los Angeles, 
and the Jet Propulsion Laboratory/California Institute of Technology, 
funded by the National Aeronautics and Space Administration.

This work made use of data supplied by the UK Swift Science Data Centre at the University of Leicester.

Funding for SDSS-III has been provided by the Alfred P. Sloan Foundation, the Participating Institutions, 
the National Science Foundation, and the U.S. Department of Energy Office of Science. 
The SDSS-III web site is http://www.sdss3.org/.
SDSS-III is managed by the Astrophysical Research Consortium for the Participating Institutions of the 
SDSS-III Collaboration including the University of Arizona, the Brazilian Participation Group,
Brookhaven National Laboratory, Carnegie Mellon University, University of Florida, the French Participation Group, 
the German Participation Group, Harvard University, the Instituto de Astrofisica de Canarias, 
the Michigan State/Notre Dame/JINA Participation Group, Johns Hopkins University,
Lawrence Berkeley National Laboratory, Max Planck Institute for Astrophysics, 
Max Planck Institute for Extraterrestrial Physics, New Mexico State University, 
New York University, Ohio State University, Pennsylvania State University, 
University of Portsmouth, Princeton University, the Spanish Participation Group,
University of Tokyo, University of Utah, Vanderbilt University, University of Virginia, 
University of Washington, and Yale University.

Part of this work is based on archival data, software or on-line services provided by the ASI Science Data Center.

This research has made use of data obtained from the high-energy Astrophysics Science Archive
Research Center (HEASARC) provided by NASA's Goddard Space Flight Center; 

The NASA/IPAC Extragalactic Database
(NED) operated by the Jet Propulsion Laboratory, California
Institute of Technology, under contract with the National Aeronautics and Space Administration.

TOPCAT\footnote{\underline{http://www.star.bris.ac.uk/$\sim$mbt/topcat/}} 
\citep{taylor05} for the preparation and manipulation of the tabular data and the images.

The work by is supported by the Programma Giovani Ricercatori - Rita Levi Montalcini - Rientro dei Cervelli (2012).
This review is also supported by the NASA grants NNX12AO97G and NNX13AP20G. 

F.M. is in debt with Dr. D'Abrusco for all the valuable discussions.

\clearpage
\bibliography{mao0404}

\begin{thebibliography}{}
\expandafter\ifx\csname natexlab\endcsname\relax\def\natexlab#1{#1}\fi

\bibitem[{{Abdo} {et~al.}(2010){Abdo}, {Ackermann}, {Agudo}, {Ajello}, {Aller},
  {Aller}, {Angelakis}, {Arkharov}, {Axelsson}, {Bach}, \& et~al.}]{abdo10}
{Abdo}, A.~A., {Ackermann}, M., {Agudo}, I., {et~al.} 2010, \apj, 716, 30

\bibitem[{{Acero} {et~al.}(2015){Acero}, {Ackermann}, {Ajello}, {Albert},
  {Atwood}, {Axelsson}, {Baldini}, {Ballet}, {Barbiellini}, {Bastieri},
  {Belfiore}, {Bellazzini}, {Bissaldi}, {Blandford}, {Bloom}, {Bogart},
  {Bonino}, {Bottacini}, {Bregeon}, {Britto}, {Bruel}, {Buehler}, {Burnett},
  {Buson}, {Caliandro}, {Cameron}, {Caputo}, {Caragiulo}, {Caraveo},
  {Casandjian}, {Cavazzuti}, {Charles}, {Chaves}, {Chekhtman}, {Cheung},
  {Chiang}, {Chiaro}, {Ciprini}, {Claus}, {Cohen-Tanugi}, {Cominsky}, {Conrad},
  {Cutini}, {D'Ammando}, {de Angelis}, {DeKlotz}, {de Palma}, {Desiante},
  {Digel}, {Di Venere}, {Drell}, {Dubois}, {Dumora}, {Favuzzi}, {Fegan},
  {Ferrara}, {Finke}, {Franckowiak}, {Fukazawa}, {Funk}, {Fusco}, {Gargano},
  {Gasparrini}, {Giebels}, {Giglietto}, {Giommi}, {Giordano}, {Giroletti},
  {Glanzman}, {Godfrey}, {Grenier}, {Grondin}, {Grove}, {Guillemot}, {Guiriec},
  {Hadasch}, {Harding}, {Hays}, {Hewitt}, {Hill}, {Horan}, {Iafrate}, {Jogler},
  {J{\'o}hannesson}, {Johnson}, {Johnson}, {Johnson}, {Johnson}, {Kamae},
  {Kataoka}, {Katsuta}, {Kuss}, {La Mura}, {Landriu}, {Larsson}, {Latronico},
  {Lemoine-Goumard}, {Li}, {Li}, {Longo}, {Loparco}, {Lott}, {Lovellette},
  {Lubrano}, {Madejski}, {Massaro}, {Mayer}, {Mazziotta}, {McEnery},
  {Michelson}, {Mirabal}, {Mizuno}, {Moiseev}, {Mongelli}, {Monzani},
  {Morselli}, {Moskalenko}, {Murgia}, {Nuss}, {Ohno}, {Ohsugi}, {Omodei},
  {Orienti}, {Orlando}, {Ormes}, {Paneque}, {Panetta}, {Perkins},
  {Pesce-Rollins}, {Piron}, {Pivato}, {Porter}, {Racusin}, {Rando}, {Razzano},
  {Razzaque}, {Reimer}, {Reimer}, {Reposeur}, {Rochester}, {Romani},
  {Salvetti}, {S{\'a}nchez-Conde}, {Saz Parkinson}, {Schulz}, {Siskind},
  {Smith}, {Spada}, {Spandre}, {Spinelli}, {Stephens}, {Strong}, {Suson},
  {Takahashi}, {Takahashi}, {Tanaka}, {Thayer}, {Thayer}, {Thompson},
  {Tibaldo}, {Tibolla}, {Torres}, {Torresi}, {Tosti}, {Troja}, {Van Klaveren},
  {Vianello}, {Winer}, {Wood}, {Wood}, {Zimmer}, \& {Fermi-LAT
  Collaboration}}]{acero15}
{Acero}, F., {Ackermann}, M., {Ajello}, M., {et~al.} 2015, \apjs, 218, 23

\bibitem[{{Ackermann} {et~al.}(2011){Ackermann}, {Ajello}, {Allafort},
  {Antolini}, {Atwood}, {Axelsson}, {Baldini}, {Ballet}, {Barbiellini},
  {Bastieri}, {Bechtol}, {Bellazzini}, {Berenji}, {Blandford}, {Bloom},
  {Bonamente}, {Borgland}, {Bottacini}, {Bouvier}, {Bregeon}, {Brigida},
  {Bruel}, {Buehler}, {Burnett}, {Buson}, {Caliandro}, {Cameron}, {Caraveo},
  {Casandjian}, {Cavazzuti}, {Cecchi}, {Charles}, {Cheung}, {Chiang},
  {Ciprini}, {Claus}, {Cohen-Tanugi}, {Conrad}, {Costamante}, {Cutini}, {de
  Angelis}, {de Palma}, {Dermer}, {Digel}, {Silva}, {Drell}, {Dubois},
  {Escande}, {Favuzzi}, {Fegan}, {Ferrara}, {Finke}, {Focke}, {Fortin},
  {Frailis}, {Fukazawa}, {Funk}, {Fusco}, {Gargano}, {Gasparrini}, {Gehrels},
  {Germani}, {Giebels}, {Giglietto}, {Giommi}, {Giordano}, {Giroletti},
  {Glanzman}, {Godfrey}, {Grenier}, {Grove}, {Guiriec}, {Gustafsson},
  {Hadasch}, {Hayashida}, {Hays}, {Healey}, {Horan}, {Hou}, {Hughes},
  {Iafrate}, {J{\'o}hannesson}, {Johnson}, {Johnson}, {Kamae}, {Katagiri},
  {Kataoka}, {Kn{\"o}dlseder}, {Kuss}, {Lande}, {Larsson}, {Latronico},
  {Longo}, {Loparco}, {Lott}, {Lovellette}, {Lubrano}, {Madejski}, {Mazziotta},
  {McConville}, {McEnery}, {Michelson}, {Mitthumsiri}, {Mizuno}, {Moiseev},
  {Monte}, {Monzani}, {Moretti}, {Morselli}, {Moskalenko}, {Murgia},
  {Nakamori}, {Naumann-Godo}, {Nolan}, {Norris}, {Nuss}, {Ohno}, {Ohsugi},
  {Okumura}, {Omodei}, {Orienti}, {Orlando}, {Ormes}, {Ozaki}, {Paneque},
  {Parent}, {Pesce-Rollins}, {Pierbattista}, {Piranomonte}, {Piron}, {Pivato},
  {Porter}, {Rain{\`o}}, {Rando}, {Razzano}, {Razzaque}, {Reimer}, {Reimer},
  {Ritz}, {Rochester}, {Romani}, {Roth}, {Sanchez}, {Sbarra}, {Scargle},
  {Schalk}, {Sgr{\`o}}, {Shaw}, {Siskind}, {Spandre}, {Spinelli}, {Strong},
  {Suson}, {Tajima}, {Takahashi}, {Takahashi}, {Tanaka}, {Thayer}, {Thayer},
  {Thompson}, {Tibaldo}, {Tinivella}, {Torres}, {Tosti}, {Troja}, {Uchiyama},
  {Vandenbroucke}, {Vasileiou}, {Vianello}, {Vitale}, {Waite}, {Wallace},
  {Wang}, {Winer}, {Wood}, {Wood}, \& {Zimmer}}]{ackermann11}
{Ackermann}, M., {Ajello}, M., {Allafort}, A., {et~al.} 2011, \apj, 743, 171

\bibitem[{{Ahn} {et~al.}(2012){Ahn}, {Alexandroff}, {Allende Prieto},
  {Anderson}, {Anderton}, {Andrews}, {Aubourg}, {Bailey}, {Balbinot}, {Barnes},
  \& et~al.}]{ahn12}
{Ahn}, C.~P., {Alexandroff}, R., {Allende Prieto}, C., {et~al.} 2012, \apjs,
  203, 21

\bibitem[{{Angel} \& {Stockman}(1980)}]{angel80}
{Angel}, J.~R.~P., \& {Stockman}, H.~S. 1980, \araa, 18, 321

\bibitem[{{Atwood} {et~al.}(2009){Atwood}, {Abdo}, {Ackermann}, {Althouse},
  {Anderson}, {Axelsson}, {Baldini}, {Ballet}, {Band}, {Barbiellini}, \&
  et~al.}]{atwood09}
{Atwood}, W.~B., {Abdo}, A.~A., {Ackermann}, M., {et~al.} 2009, \apj, 697, 1071

\bibitem[{{Becker} {et~al.}(1995){Becker}, {White}, \& {Helfand}}]{becker95}
{Becker}, R.~H., {White}, R.~L., \& {Helfand}, D.~J. 1995, \apj, 450, 559

\bibitem[{{Blandford} \& {Rees}(1978)}]{blanford78}
{Blandford}, R.~D., \& {Rees}, M.~J. 1978, \physscr, 17, 265

\bibitem[{{Burrows} {et~al.}(2005){Burrows}, {Hill}, {Nousek}, {Kennea},
  {Wells}, {Osborne}, {Abbey}, {Beardmore}, {Mukerjee}, {Short}, {Chincarini},
  {Campana}, {Citterio}, {Moretti}, {Pagani}, {Tagliaferri}, {Giommi},
  {Capalbi}, {Tamburelli}, {Angelini}, {Cusumano}, {Br{\"a}uninger}, {Burkert},
  \& {Hartner}}]{burrows05}
{Burrows}, D.~N., {Hill}, J.~E., {Nousek}, J.~A., {et~al.} 2005, \ssr, 120, 165

\bibitem[{{Chen} \& {Bai}(2011)}]{chen11}
{Chen}, L., \& {Bai}, J.~M. 2011, \apj, 735, 108

\bibitem[{{Condon} {et~al.}(1998){Condon}, {Cotton}, {Greisen}, {Yin},
  {Perley}, {Taylor}, \& {Broderick}}]{condon98}
{Condon}, J.~J., {Cotton}, W.~D., {Greisen}, E.~W., {et~al.} 1998, \aj, 115,
  1693

\bibitem[{{D'Abrusco} {et~al.}(2013){D'Abrusco}, {Massaro}, {Paggi}, {Masetti},
  {Tosti}, {Giroletti}, \& {Smith}}]{dabrusco13}
{D'Abrusco}, R., {Massaro}, F., {Paggi}, A., {et~al.} 2013, \apjs, 206, 12

\bibitem[{{D'Abrusco} {et~al.}(2014){D'Abrusco}, {Massaro}, {Paggi}, {Smith},
  {Masetti}, {Landoni}, \& {Tosti}}]{dabrusco14}
---. 2014, \apjs, 215, 14

\bibitem[{{Donato} {et~al.}(2001){Donato}, {Ghisellini}, {Tagliaferri}, \&
  {Fossati}}]{donato01}
{Donato}, D., {Ghisellini}, G., {Tagliaferri}, G., \& {Fossati}, G. 2001, \aap,
  375, 739

\bibitem[{{Draine}(2003)}]{draine03}
{Draine}, B.~T. 2003, \araa, 41, 241

\bibitem[{{Evans} {et~al.}(2014){Evans}, {Osborne}, {Beardmore}, {Page},
  {Willingale}, {Mountford}, {Pagani}, {Burrows}, {Kennea}, {Perri},
  {Tagliaferri}, \& {Gehrels}}]{evans14}
{Evans}, P.~A., {Osborne}, J.~P., {Beardmore}, A.~P., {et~al.} 2014, \apjs,
  210, 8

\bibitem[{{Feigelson} \& {Jogesh Babu}(2012)}]{feigelson12}
{Feigelson}, E.~D., \& {Jogesh Babu}, G. 2012, {Modern Statistical Methods for
  Astronomy}

\bibitem[{{Fichtel} {et~al.}(1994){Fichtel}, {Bertsch}, {Chiang}, {Dingus},
  {Esposito}, {Fierro}, {Hartman}, {Hunter}, {Kanbach}, {Kniffen}, {Kwok},
  {Lin}, {Mattox}, {Mayer-Hasselwander}, {McDonald}, {Michelson}, {von
  Montigny}, {Nolan}, {Pinkau}, {Radecke}, {Rothermel}, {Sreekumar}, {Sommer},
  {Schneid}, {Thompson}, \& {Willis}}]{fichtel94}
{Fichtel}, C.~E., {Bertsch}, D.~L., {Chiang}, J., {et~al.} 1994, \apjs, 94, 551

\bibitem[{{Fossati} {et~al.}(1997){Fossati}, {Celotti}, {Ghisellini}, \&
  {Maraschi}}]{fossati97}
{Fossati}, G., {Celotti}, A., {Ghisellini}, G., \& {Maraschi}, L. 1997, \mnras,
  289, 136

\bibitem[{{Fossati} {et~al.}(1998){Fossati}, {Maraschi}, {Celotti}, {Comastri},
  \& {Ghisellini}}]{fossati98}
{Fossati}, G., {Maraschi}, L., {Celotti}, A., {Comastri}, A., \& {Ghisellini},
  G. 1998, \mnras, 299, 433

\bibitem[{{Ghisellini} {et~al.}(1998){Ghisellini}, {Celotti}, {Fossati},
  {Maraschi}, \& {Comastri}}]{ghisellini98}
{Ghisellini}, G., {Celotti}, A., {Fossati}, G., {Maraschi}, L., \& {Comastri},
  A. 1998, \mnras, 301, 451

\bibitem[{{Ghisellini} {et~al.}(2009){Ghisellini}, {Maraschi}, \&
  {Tavecchio}}]{ghisellini09}
{Ghisellini}, G., {Maraschi}, L., \& {Tavecchio}, F. 2009, \mnras, 396, L105

\bibitem[{{Ghisellini} \& {Tavecchio}(2008)}]{ghisellini08}
{Ghisellini}, G., \& {Tavecchio}, F. 2008, \mnras, 387, 1669

\bibitem[{{Ghisellini} {et~al.}(2012){Ghisellini}, {Tavecchio}, {Foschini},
  {Sbarrato}, {Ghirlanda}, \& {Maraschi}}]{ghisellini12}
{Ghisellini}, G., {Tavecchio}, F., {Foschini}, L., {et~al.} 2012, \mnras, 425,
  1371

\bibitem[{{Ghisellini} {et~al.}(2011){Ghisellini}, {Tagliaferri}, {Foschini},
  {Ghirlanda}, {Tavecchio}, {Della Ceca}, {Haardt}, {Volonteri}, \&
  {Gehrels}}]{ghisellini11}
{Ghisellini}, G., {Tagliaferri}, G., {Foschini}, L., {et~al.} 2011, \mnras,
  411, 901

\bibitem[{{Giommi} {et~al.}(1999){Giommi}, {Menna}, \& {Padovani}}]{giommi99}
{Giommi}, P., {Menna}, M.~T., \& {Padovani}, P. 1999, \mnras, 310, 465

\bibitem[{{Giommi} \& {Padovani}(1994)}]{giommi94}
{Giommi}, P., \& {Padovani}, P. 1994, \mnras, 268, L51

\bibitem[{{Giommi} {et~al.}(2013){Giommi}, {Padovani}, \& {Polenta}}]{giommi13}
{Giommi}, P., {Padovani}, P., \& {Polenta}, G. 2013, \mnras, 431, 1914

\bibitem[{{Giommi} {et~al.}(2012){Giommi}, {Padovani}, {Polenta}, {Turriziani},
  {D'Elia}, \& {Piranomonte}}]{giommi12}
{Giommi}, P., {Padovani}, P., {Polenta}, G., {et~al.} 2012, \mnras, 420, 2899

\bibitem[{{Hinshaw} {et~al.}(2013){Hinshaw}, {Larson}, {Komatsu}, {Spergel},
  {Bennett}, {Dunkley}, {Nolta}, {Halpern}, {Hill}, {Odegard}, {Page}, {Smith},
  {Weiland}, {Gold}, {Jarosik}, {Kogut}, {Limon}, {Meyer}, {Tucker}, {Wollack},
  \& {Wright}}]{hinshaw13}
{Hinshaw}, G., {Larson}, D., {Komatsu}, E., {et~al.} 2013, \apjs, 208, 19

\bibitem[{{Kalberla} {et~al.}(2005){Kalberla}, {Burton}, {Hartmann}, {Arnal},
  {Bajaja}, {Morras}, \& {P{\"o}ppel}}]{kalberla05}
{Kalberla}, P.~M.~W., {Burton}, W.~B., {Hartmann}, D., {et~al.} 2005, \aap,
  440, 775

\bibitem[{{Landt} {et~al.}(2001){Landt}, {Padovani}, {Perlman}, {Giommi},
  {Bignall}, \& {Tzioumis}}]{landt01}
{Landt}, H., {Padovani}, P., {Perlman}, E.~S., {et~al.} 2001, \mnras, 323, 757

\bibitem[{{Maraschi} {et~al.}(1986){Maraschi}, {Ghisellini}, {Tanzi}, \&
  {Treves}}]{maraschi86}
{Maraschi}, L., {Ghisellini}, G., {Tanzi}, E.~G., \& {Treves}, A. 1986, \apj,
  310, 325

\bibitem[{{Massaro} {et~al.}(2011{\natexlab{a}}){Massaro}, {Giommi}, {Leto},
  {Marchegiani}, {Maselli}, {Perri}, \& {Piranomonte}}]{massaro11}
{Massaro}, E., {Giommi}, P., {Leto}, C., {et~al.} 2011{\natexlab{a}},
  {Multifrequency Catalogue of Blazars (3rd Edition)}

\bibitem[{{Massaro} {et~al.}(2009){Massaro}, {Giommi}, {Leto}, {Marchegiani},
  {Maselli}, {Perri}, {Piranomonte}, \& {Sclavi}}]{massaro09}
---. 2009, \aap, 495, 691

\bibitem[{{Massaro} {et~al.}(2014{\natexlab{a}}){Massaro}, {Maselli}, {Leto},
  {Marchegiani}, {Perri}, {Giommi}, \& {Piranomonte}}]{massaro15}
{Massaro}, E., {Maselli}, A., {Leto}, C., {et~al.} 2014{\natexlab{a}},
  {Multifrequency Catalogue of Blazars - 5th Edition}

\bibitem[{{Massaro} {et~al.}(2012){Massaro}, {Nesci}, \&
  {Piranomonte}}]{massaro12}
{Massaro}, E., {Nesci}, R., \& {Piranomonte}, S. 2012, \mnras, 422, 2322

\bibitem[{{Massaro} {et~al.}(2013){Massaro}, {D'Abrusco}, {Giroletti}, {Paggi},
  {Masetti}, {Tosti}, {Nori}, \& {Funk}}]{massarof13}
{Massaro}, F., {D'Abrusco}, R., {Giroletti}, M., {et~al.} 2013, \apjs, 207, 4

\bibitem[{{Massaro} {et~al.}(2014{\natexlab{b}}){Massaro}, {Masetti},
  {D'Abrusco}, {Paggi}, \& {Funk}}]{massarof14}
{Massaro}, F., {Masetti}, N., {D'Abrusco}, R., {Paggi}, A., \& {Funk}, S.
  2014{\natexlab{b}}, \aj, 148, 66

\bibitem[{{Massaro} {et~al.}(2011{\natexlab{b}}){Massaro}, {Paggi}, {Elvis}, \&
  {Cavaliere}}]{massarof11}
{Massaro}, F., {Paggi}, A., {Elvis}, M., \& {Cavaliere}, A. 2011{\natexlab{b}},
  \apj, 739, 73

\bibitem[{{Massaro} {et~al.}(2008){Massaro}, {Tramacere}, {Cavaliere}, {Perri},
  \& {Giommi}}]{massarof08}
{Massaro}, F., {Tramacere}, A., {Cavaliere}, A., {Perri}, M., \& {Giommi}, P.
  2008, \aap, 478, 395

\bibitem[{{Mauch} {et~al.}(2003){Mauch}, {Murphy}, {Buttery}, {Curran},
  {Hunstead}, {Piestrzynski}, {Robertson}, \& {Sadler}}]{mauch03}
{Mauch}, T., {Murphy}, T., {Buttery}, H.~J., {et~al.} 2003, \mnras, 342, 1117

\bibitem[{{Meyer} {et~al.}(2011){Meyer}, {Fossati}, {Georganopoulos}, \&
  {Lister}}]{meyer11}
{Meyer}, E.~T., {Fossati}, G., {Georganopoulos}, M., \& {Lister}, M.~L. 2011,
  \apj, 740, 98

\bibitem[{{Mukai}(1993)}]{mukai93}
{Mukai}, K. 1993, Legacy, vol.~3, p.21-31, 3, 21

\bibitem[{{Nieppola} {et~al.}(2006){Nieppola}, {Tornikoski}, \&
  {Valtaoja}}]{nieppola06}
{Nieppola}, E., {Tornikoski}, M., \& {Valtaoja}, E. 2006, \aap, 445, 441

\bibitem[{{Nieppola} {et~al.}(2008){Nieppola}, {Valtaoja}, {Tornikoski},
  {Hovatta}, \& {Kotiranta}}]{nieppola08}
{Nieppola}, E., {Valtaoja}, E., {Tornikoski}, M., {Hovatta}, T., \&
  {Kotiranta}, M. 2008, \aap, 488, 867

\bibitem[{{Padovani}(2007)}]{padovani07}
{Padovani}, P. 2007, \apss, 309, 63

\bibitem[{{Pavlidou} {et~al.}(2012){Pavlidou}, {Richards}, {Max-Moerbeck},
  {King}, {Pearson}, {Readhead}, {Reeves}, {Stevenson}, {Angelakis},
  {Fuhrmann}, {Zensus}, {Giroletti}, {Reimer}, {Healey}, {Romani}, \&
  {Shaw}}]{pavlidou12}
{Pavlidou}, V., {Richards}, J.~L., {Max-Moerbeck}, W., {et~al.} 2012, \apj,
  751, 149

\bibitem[{{Plotkin} {et~al.}(2012){Plotkin}, {Anderson}, {Brandt}, {Markoff},
  {Shemmer}, \& {Wu}}]{plotkin12}
{Plotkin}, R.~M., {Anderson}, S.~F., {Brandt}, W.~N., {et~al.} 2012, \apjl,
  745, L27

\bibitem[{{Sambruna} {et~al.}(1996){Sambruna}, {Maraschi}, \&
  {Urry}}]{sambruna96}
{Sambruna}, R.~M., {Maraschi}, L., \& {Urry}, C.~M. 1996, \apj, 463, 444

\bibitem[{{Shields}(1978)}]{shields78}
{Shields}, G.~A. 1978, \nat, 272, 706

\bibitem[{{Stocke} {et~al.}(1985){Stocke}, {Liebert}, {Schmidt}, {Gioia},
  {Maccacaro}, {Schild}, {Maccagni}, \& {Arp}}]{stocke85}
{Stocke}, J.~T., {Liebert}, J., {Schmidt}, G., {et~al.} 1985, \apj, 298, 619

\bibitem[{{Stoughton} {et~al.}(2002){Stoughton}, {Lupton}, {Bernardi},
  {Blanton}, {Burles}, {Castander}, {Connolly}, {Eisenstein}, {Frieman},
  {Hennessy}, {Hindsley}, {Ivezi{\'c}}, {Kent}, {Kunszt}, {Lee}, {Meiksin},
  {Munn}, {Newberg}, {Nichol}, {Nicinski}, {Pier}, {Richards}, {Richmond},
  {Schlegel}, {Smith}, {Strauss}, {SubbaRao}, {Szalay}, {Thakar}, {Tucker},
  {Vanden Berk}, {Yanny}, {Adelman}, {Anderson}, {Anderson}, {Annis},
  {Bahcall}, {Bakken}, {Bartelmann}, {Bastian}, {Bauer}, {Berman},
  {B{\"o}hringer}, {Boroski}, {Bracker}, {Briegel}, {Briggs}, {Brinkmann},
  {Brunner}, {Carey}, {Carr}, {Chen}, {Christian}, {Colestock}, {Crocker},
  {Csabai}, {Czarapata}, {Dalcanton}, {Davidsen}, {Davis}, {Dehnen},
  {Dodelson}, {Doi}, {Dombeck}, {Donahue}, {Ellman}, {Elms}, {Evans}, {Eyer},
  {Fan}, {Federwitz}, {Friedman}, {Fukugita}, {Gal}, {Gillespie}, {Glazebrook},
  {Gray}, {Grebel}, {Greenawalt}, {Greene}, {Gunn}, {de Haas}, {Haiman},
  {Haldeman}, {Hall}, {Hamabe}, {Hansen}, {Harris}, {Harris}, {Harvanek},
  {Hawley}, {Hayes}, {Heckman}, {Helmi}, {Henden}, {Hogan}, {Hogg}, {Holmgren},
  {Holtzman}, {Huang}, {Hull}, {Ichikawa}, {Ichikawa}, {Johnston}, {Kauffmann},
  {Kim}, {Kimball}, {Kinney}, {Klaene}, {Kleinman}, {Klypin}, {Knapp},
  {Korienek}, {Krolik}, {Kron}, {Krzesi{\'n}ski}, {Lamb}, {Leger},
  {Limmongkol}, {Lindenmeyer}, {Long}, {Loomis}, {Loveday}, {MacKinnon},
  {Mannery}, {Mantsch}, {Margon}, {McGehee}, {McKay}, {McLean}, {Menou},
  {Merelli}, {Mo}, {Monet}, {Nakamura}, {Narayanan}, {Nash}, {Neilsen},
  {Newman}, {Nitta}, {Odenkirchen}, {Okada}, {Okamura}, {Ostriker}, {Owen},
  {Pauls}, {Peoples}, {Peterson}, {Petravick}, {Pope}, {Pordes}, {Postman},
  {Prosapio}, {Quinn}, {Rechenmacher}, {Rivetta}, {Rix}, {Rockosi}, {Rosner},
  {Ruthmansdorfer}, {Sandford}, {Schneider}, {Scranton}, {Sekiguchi}, {Sergey},
  {Sheth}, {Shimasaku}, {Smee}, {Snedden}, {Stebbins}, {Stubbs}, {Szapudi},
  {Szkody}, {Szokoly}, {Tabachnik}, {Tsvetanov}, {Uomoto}, {Vogeley}, {Voges},
  {Waddell}, {Walterbos}, {Wang}, {Watanabe}, {Weinberg}, {White}, {White},
  {Wilhite}, {Wolfe}, {Yasuda}, {York}, {Zehavi}, \& {Zheng}}]{stoughton02}
{Stoughton}, C., {Lupton}, R.~H., {Bernardi}, M., {et~al.} 2002, \aj, 123, 485

\bibitem[{{Taylor}(2005)}]{taylor05}
{Taylor}, M.~B. 2005, in Astronomical Society of the Pacific Conference Series,
  Vol. 347, Astronomical Data Analysis Software and Systems XIV, ed.
  P.~{Shopbell}, M.~{Britton}, \& R.~{Ebert}, 29

\bibitem[{{Urry} \& {Mushotzky}(1982)}]{urry82}
{Urry}, C.~M., \& {Mushotzky}, R.~F. 1982, \apj, 253, 38

\bibitem[{{Urry} \& {Padovani}(1995)}]{urry95}
{Urry}, C.~M., \& {Padovani}, P. 1995, \pasp, 107, 803

\bibitem[{{White} {et~al.}(1997){White}, {Becker}, {Helfand}, \&
  {Gregg}}]{white97}
{White}, R.~L., {Becker}, R.~H., {Helfand}, D.~J., \& {Gregg}, M.~D. 1997,
  \apj, 475, 479

\bibitem[{{Wright} {et~al.}(2010){Wright}, {Eisenhardt}, {Mainzer}, {Ressler},
  {Cutri}, {Jarrett}, {Kirkpatrick}, {Padgett}, {McMillan}, {Skrutskie},
  {Stanford}, {Cohen}, {Walker}, {Mather}, {Leisawitz}, {Gautier}, {McLean},
  {Benford}, {Lonsdale}, {Blain}, {Mendez}, {Irace}, {Duval}, {Liu}, {Royer},
  {Heinrichsen}, {Howard}, {Shannon}, {Kendall}, {Walsh}, {Larsen}, {Cardon},
  {Schick}, {Schwalm}, {Abid}, {Fabinsky}, {Naes}, \& {Tsai}}]{wright10}
{Wright}, E.~L., {Eisenhardt}, P.~R.~M., {Mainzer}, A.~K., {et~al.} 2010, \aj,
  140, 1868

\end{thebibliography}
\clearpage

\end{document}